\documentclass[11pt]{article}
\usepackage[utf8]{inputenc}
\usepackage{jheppub}
\usepackage{amsmath,amssymb,amsfonts,graphicx,slashed,color,amsthm,mathtools,upgreek, enumerate, tensor, scalerel, subfig}
\usepackage{arydshln}
\usepackage[dvipsnames]{xcolor}
\usepackage{bbold}
\allowdisplaybreaks

\makeatletter
\gdef\@fpheader{}
\makeatother

\newcommand{\bx}{\mathbf{x}}
\newcommand{\ka}{\alpha}
\newcommand{\beq}{\begin{equation}}
\newcommand{\eeq}{\end{equation}}
\newcommand{\beqn}{\begin{eqnarray}}
\newcommand{\eeqn}{\end{eqnarray}}
\newcommand{\pa}{\partial}

\renewcommand{\d}{\mathrm{d}}

\newcommand{\cE}{\mathcal{E}}
\newcommand{\cF}{\mathcal{F}}
\newcommand{\en}{\text{env}}

\newcommand{\ph}{\text{phys}}
\newcommand{\co}{\text{code}}
\newcommand{\re}{\text{ref}}
\newcommand{\cH}{\mathcal{H}}

\newcommand{\bra}[1]{\langle #1 \vert}
\newcommand{\ket}[1]{\vert #1 \rangle}
\newcommand{\inner}[2]{\langle #1 \vert #2 \rangle}
\newcommand{\avg}[1]{\langle #1 \rangle}
\DeclareMathOperator{\Tr}{Tr}
\DeclareMathOperator{\Pf}{Pf}
\newcommand{\cR}{\mathcal{R}}

\newcommand{\Abar}{{\overline{A}}}

\newcommand{\rerr}[1]{\mathcal{R}\circ\mathcal{E}(#1)}

\def\stretchint#1{\vcenter{\hbox{\stretchto[440]{\displaystyle\int}{#1}}}}

\DeclareMathOperator{\sech}{sech}

\title{Quantum Error Correction from Complexity in Brownian SYK}
\author[a,b]{Vijay Balasubramanian}
\author[c]{\!, Arjun Kar}
\author[a]{\!, Cathy Li}
\author[d]{\!, Onkar Parrikar} 
\author[d]{\!, Harshit Rajgadia} 

\affiliation[\,a]{David Rittenhouse Laboratory, University of Pennsylvania,\\
209 S. 33rd Street, Philadelphia PA 19104, USA.}
\affiliation[\,b]{Theoretische Natuurkunde, Vrije Universiteit Brussel (VUB), and \\ International Solvay Institutes, Pleinlaan 2, B-1050 Brussels, Belgium.}
\affiliation[\,c]{Department of Physics and Astronomy, University of British Columbia, \\
6224 Agricultural Road, Vancouver, BC V6T 1Z1, Canada.}
\affiliation[\,d]{Department of Theoretical Physics,
Tata Institute for Fundamental Research,\\ Mumbai 400005, India.}

\emailAdd{vijay@physics.upenn.edu}
\emailAdd{arjunkar@phas.ubc.ca}
\emailAdd{yl244@sas.upenn.edu}
\emailAdd{parrikar@theory.tifr.res.in}
\emailAdd{hrajgadia@gmail.com}

\begin{document}

\abstract{
 We study the robustness of quantum error correction in a one-parameter ensemble of codes generated by the Brownian SYK model, where the parameter quantifies the encoding complexity. The robustness of error correction by a quantum code is upper bounded by the ``mutual purity'' of a certain entangled state
 between the code subspace and environment in the isometric extension of the error channel, where the mutual purity of a density matrix $\rho_{AB}$ is the difference $\cF_\rho (A:B) \equiv \Tr \rho_{AB}^2 - \Tr \rho_A^2 \Tr \rho_B^2$.
 We show that when the encoding complexity is small, the mutual purity is $O(1)$ for  the erasure of a small number of qubits (i.e., the encoding is fragile).  However, this quantity decays exponentially,  becoming $O(1/N)$ for $O(\log N)$ encoding complexity. Further, at polynomial encoding complexity, the mutual purity saturates to a plateau of $O(e^{-N})$.  We also find a hierarchy of complexity scales associated to a tower of subleading contributions to the mutual purity that quantitatively, but not qualitatively, adjust our error correction bound as encoding complexity increases.  In the AdS/CFT context, our results suggest that any portion of the entanglement wedge of a general boundary subregion $A$ with sufficiently high encoding complexity is robustly protected against low-rank errors acting on $A$ with no prior access to the encoding map. From the bulk point of view, we expect such bulk degrees of freedom to be causally inaccessible from the region $A$ despite being encoded in it.
 }
\maketitle

\parskip=12pt

\section{Introduction}

The bulk-to-boundary map in AdS/CFT has a rich structure.  For any boundary subregion $A$, the associated Ryu-Takayanagi surface \cite{Ryu:2006bv} singles out a certain subregion $a$ of the bulk spacetime called the entanglement wedge of $A$ \cite{Headrick:2014cta}. The AdS/CFT map then satisfies subregion duality: bulk semi-classical degrees of freedom in $a$ are encoded within $A$ and are protected against erasures in $\overline{A}$. Furthermore, bulk operators within the entanglement wedge $a$ can be reconstructed as boundary operators localized within the boundary region $A$, a property sometimes known as entanglement wedge reconstruction \cite{Dong:2016eik}. 
Using the language of \cite{Verlinde:2012cy, Papadodimas:2013jku, Lewkowycz:2013nqa,Jafferis:2015del,Almheiri:2014lwa,Engelhardt:2014gca,Dong:2016eik,Penington:2019kki}, these properties hold because the Ryu-Takayanagi formula and its quantum generalizations imply that the bulk-to-boundary map in AdS/CFT is a quantum error correcting code with complementary recovery, where the entanglement wedge $a$ of $A$ is protected against the erasure of $\overline{A}$, while $\overline{a}$ is protected against the erasure of $A$.  
\begin{figure}
    \centering
    \includegraphics[height=4cm]{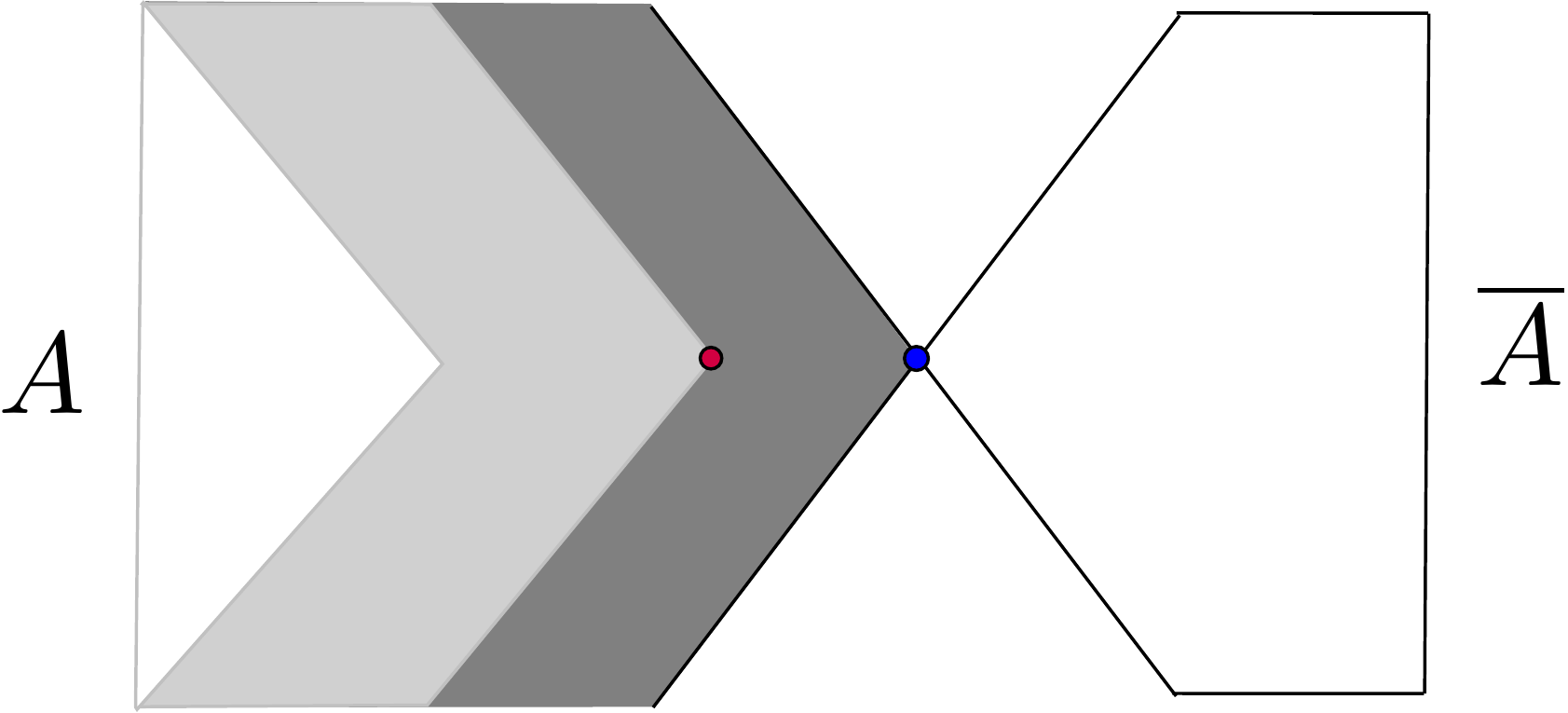}
    \caption{\small{The entanglement wedge of a boundary subregion $A$ can have a rich substructure. The outer white region on the left is the causal wedge of $A$, while the slightly darker grey region bounded by a non-minimal QES (shown in red) is the simple wedge. Beyond this lies the python's lunch (dark grey).}}
    \label{fig:PL1}
\end{figure}

Recent progress points towards a sharper characterization of the structure of entanglement wedges that appear in holography and its generalizations \cite{Almheiri:2019psf,Penington:2019npb,  Almheiri:2019hni, Engelhardt:2021qjs,Engelhardt:2021mue,Akers:2021fut}.
Given a general boundary subregion, the corresponding entanglement wedge has a layered structure, i.e., it can be broken up into three regions: the causal wedge, the simple wedge \cite{Engelhardt:2021qjs, Engelhardt:2021mue} and the python's lunch \cite{Brown:2019rox}.
These three regions of the entanglement wedge are defined as follows (see Figure~\ref{fig:PL1}):
the causal wedge is the region in the bulk which is causally accessible from the boundary, i.e., a boundary observer in the domain of dependence of $A$ can send signals to and receive signals from all points in the causal wedge of $A$.
The simple wedge is defined to be the bulk domain of dependence of the homology region between $A$ and the outermost quantum extremal surface (QES) which need not be minimal among all the QESs associated with $A$. The simple wedge is generically larger than the causal wedge, and so there are points in the simple wedge which are out of causal contact with the domain of dependence of $A$. However, it has been argued that the simple wedge can always be brought in causal contact with the boundary by performing backwards and forwards Lorentzian time evolution with sources turned on to de-focus the causal horizons \cite{Engelhardt:2021mue}. Finally, the python's lunch region is defined as the portion of the entanglement wedge which lies between the outermost QES and the minimal QES. This region is causally inaccessible from the boundary subregion $A$, and furthermore since it lies behind an extremal surface it cannot be brought into causal contact with the domain of dependence of $A$ (in contrast with the simple wedge); this follows from the fact that extremal surfaces must always lie behind causal horizons. This seems to lead to a puzzle --- on the one hand, bulk operators in the python's lunch are encoded in $A$ and in particular one should be able to create a semi-classical bulk excitation in the python's lunch via an operator acting on the domain of dependence of $A$. On the other hand, semi-classical gravity seems to forbid this!

The evaporating black hole provides a context where this apparent contradiction is particularly sharp. Beyond the Page time, a portion of the black hole interior --- the island --- lies in the entanglement wedge of the radiation (see  \cite{Almheiri:2019psf,Penington:2019npb,Almheiri:2019hni,Almheiri:2019qdq,Balasubramanian:2020hfs, Geng:2021hlu, Geng:2021wcq, Geng:2020fxl,Balasubramanian:2020coy,Hartman:2020khs,Anderson:2020vwi,Balasubramanian:2021wgd} for a partial list of articles discussing this phenomenon for black hole and cosmological horizons).
But this portion lies behind a non-minimal QES, namely the empty surface,\footnote{In more detail, in the toy models where these calculations are possible, the radiation is extracted into an auxiliary reservoir that is not geometrically connected to the island. Even in the absence of a geometric connection, there is still an obvious candidate extremal surface which one can consider as bounding the region dual to the radiation, namely the ``empty surface". By this, we mean that the entire black hole spacetime is taken to lie ``outside" the would-be entanglement wedge. After the Page time, this surface is no longer the quantum minimal surface which computes the radiation entropy, and the true minimum QES  lies in the spacetime near the black hole horizon. Nevertheless, the island is  ``behind the empty extremal surface" from the point of view of the radiation.
}
and therefore constitutes a python's lunch. While one should be able to manipulate operators in the island by quantum operations on the radiation, such operations seem to blatantly violate semi-classical bulk causality. A potential way out is suggested by bounds coming from computational complexity \cite{Harlow:2013tf, Brown:2019rox,Zhao:2020wgp,Engelhardt:2021mue,Engelhardt:2021qjs,Akers:2022qdl,Kim:2020cds,Balasubramanian:2022fiy,Kar:2022qkf} --- we expect that the encoding map for excitations in the python's lunch region is extremely complex, perhaps exponentially so in the number of qubits, and so any computationally bounded observer (with access only to sub-exponential operations on the radiation) will be unable to manipulate the degrees of freedom in the island. This is how we expect that semi-classical bulk causality will be approximately respected. On the other hand, certain finely tuned, exponentially complex operations on the radiation should be able to manipulate degrees of freedom in the island, but the gravitational mechanism for this involves Euclidean wormholes.

We can get an intuition for why complexity can protect information in this way from an analogy to older results concerning the complexity of black hole microstates and the difficulty of using simple probes to extract information about them \cite{Balasubramanian:2005kk,Balasubramanian:2005mg,Balasubramanian:2005qu}.  Consider, for example, a Schwarzschild black hole of mass $M$ in AdS$_5$ with a length scale $\ell$.   A microstate of this black hole is described in the dual $SU(N)$ Yang-Mills theory with 16 supersymmetries, by an operator ${\cal O}$ of dimension $\Delta= M \ell \sim N^2$.   ${\cal O}$ is roughly a polynomial of length $N^2$ built from  the elementary fields of the Yang-Mills theory (a gauge field $A_\mu$, fermions $\psi_a$ and three complex adjoint scalars $X,Y,Z$) and their derivatives,  with indices contracted to make the polynomial gauge and Lorentz invariant.  Almost all such long polynomials are random sequences of fields and derivatives up to constraints of gauge and Lorentz invariance.  A light probe of the state like the graviton corresponds to an operator of dimension $O(1)$, like $P={\rm Tr}(XX)$. The question is whether a measurement, modeled as a correlation function in the state created by ${\cal O}$, $\langle 0| {\cal O}^\dagger P^\dagger P {\cal O} |0\rangle$, can reveal information about the identity of ${\cal O}$.  The authors of \cite{Balasubramanian:2005kk,Balasubramanian:2005mg} argue that the answer is ``no'' because of the universal statistics of random polynomials, which mean that almost all ${\cal O}$ will lead to a similar sum of terms from contractions between the fields in the probe and the fields in ${\cal O}$ in evaluating the correlator.  As such, simple (i.e., low-dimension) probes cannot reveal the microstate, but an observer with prior knowledge of the state could construct a fine-tuned, complex probe to check that knowledge, by choosing these probes to match long sequences of the fields composing ${\cal O}$. One expects the situation in the python's lunch inside an evaporating black hole to be somewhat analogous: a highly complex encoding map prevents simple operations in the radiation from affecting the black hole interior, but if the encoding map is accessible then finely tuned, complex operations affecting the interior may be performed more easily.

Kim, Preskill and Tang (KPT) have sharpened these  expectations \cite{Kim:2020cds}.
They suggested that the encoding of the black hole interior degrees of freedom in the radiation, thought of as a quantum error correcting code, has robust error correction properties against low-rank, computationally bounded errors on the radiation, or more precisely, errors which effectively see the radiation density matrix as thermal. Note that this is not the standard error correction one encounters in the context of subregion duality; in the KPT formulation, bulk degrees of freedom in the island --- while being encoded in the radiation --- are nevertheless approximately (up to corrections exponentially small in the black hole entropy) protected against certain errors acting on the radiation itself. KPT then argued that this approximate error correction implies the existence of ``ghost logical'' operators which act on the radiation to mimic bulk operators in the island and at the same time commute with computationally bounded operators on the radiation --- thus realizing the approximate causality of the black hole spacetime. The language of quantum error correction thus enables one to formulate and address the question of bulk causality in a universal manner. 

Recently, the novel error correction in evaporating black holes proposed by KPT was tested in  a toy model for an evaporating black hole in Jackiw-Teitelboim gravity \cite{Balasubramanian:2022fiy}, and it was argued that the bulk degrees of freedom in the island are protected against a large class of low-rank error operations on the radiation which do not have access to the details of the microscopic black hole state. The low-rank criterion can be formalized as a bound on the coherent information of the error in terms of the black hole entropy. In \cite{Balasubramanian:2022fiy}, it was also conjectured that this same robust error correction should also work in the python's lunch portions of more general entanglement wedges. The underlying reason  is the high complexity of encoding in the python's lunch. The rough picture is the same as KPT -- as the encoding map becomes sufficiently complex, any generic, low-rank error operation involving ``simple'' operations sees only a coarse-grained\footnote{The relevant notion of coarse-graining was defined in \cite{Engelhardt:2018kcs, Engelhardt:2021mue}: one finds the maximum-entropy state consistent with correlation functions of all simple operators, including Lorentizan time-folds with simple sources turned on. Here simple operators and sources are defined as those whose effects propagate causally in the bulk.} density matrix on the boundary subregion, with no sign of the encoded subspace. In other words, the encoded subspace gets lost within the exponentially large Hilbert space of the boundary subregion. This ``complexity-protected error correction''  makes it possible for the semi-classical degrees of freedom in the python's lunch to be encoded in a boundary subregion and yet be causally inaccessible from it using simple probes. 

The purpose of this paper is to demonstrate the above phenomenon in a toy model where the behavior of the encoding complexity is known more or less by construction. 
Such control is difficult to achieve directly in real holography because proving results about the complexity of the bulk-to-boundary map (without resorting to toy models like tensor networks) in different regions of the entanglement wedge is generically a very difficult task.
Rather than a single code, here we consider an ensemble of quantum error correcting codes of the type relevant for entanglement wedge reconstruction in AdS/CFT. Since we want control over the complexity of encoding, our ensemble of codes is generated by picking the encoding map from an ensemble of unitaries with fixed circuit complexity. 

We accomplish this by taking these unitaries to be time evolution operators $U(T)= \mathcal{T}\exp [-i\int_0^T\d t\; H(t) ]$ in the Brownian Sachdev-Ye-Kitaev (SYK) model \cite{Kitaev2015v1,Kitaev2015v2,Maldacena:2016hyu,Saad:2018bqo,Stanford:2021bhl}, a quantum mechanical theory of $N$ Majorana fermions. The SYK model here is merely a trick to generate an ensemble of unitaries with fixed complexity, parametrized by the number $T$. When  $T$ is small, the corresponding set of unitary operators is clustered around the identity operator, but as $T \to \infty$ this set grows \cite{Brown:2017jil, Brandao:2019sgy, Balasubramanian:2019wgd, Haferkamp:2021uxo, Balasubramanian:2021mxo} to cover (modulo global symmetries) the entire unitary group.
When it covers the entire unitary group, the typical complexity of an operator in the set is exponentially large \cite{Susskind:2018pmk}.
So, computing the average error correction properties of such sets gives us some insight into the behavior of a family of codes with increasing complexity.

In this paper we will consider typical, low-rank errors with no prior access to the encoding map, and acting on a small fixed fraction of the physical Hilbert space. As a particular instance of such errors, we will consider the erasure of a small fraction of the physical Hilbert space. In quantum information theory, it is standard to model an error in terms of coupling to an external environment and tracing out the environment. The error correction properties of the code can then be studied in terms of the amount of correlation generated by the error between the code subspace and the environment. Error correction works with high accuracy when these correlations are suppressed by a large parameter e.g. the dimension of the physical Hilbert space.

We will study a particular measure of correlation, namely the ``mutual purity'' between the code subspace and the environment. 
We define the mutual purity $\cF_\rho (A:B)$ of a density matrix $\rho$ between Hilbert subsystems $\cH_A$ and $\cH_B$ as $\Tr \rho_{AB}^2 - \Tr \rho_A^2 \Tr \rho_B^2$.
The fact that this quantity is a good measure of error correction is rigorously justified in Appendix \ref{appQEC}. Our main result is that for the Brownian SYK ensemble of quantum error correcting codes, there are three complexity regimes of interest. 
\begin{enumerate}[(i)]
\item For $T$ smaller than a scrambling time $T \sim \log N$ (i.e., low encoding complexity) the erasure of a small fraction of the physical qubits generate an $O(1)$ amount of correlation that  decays exponentially with $T$ between the code subspace and the environment, and thus there is no robust quantum error correction.
\item For $T> \log N$, the mutual purity becomes $O(1/N)$  but keeps decaying further as the complexity $T$ increases. \item When $T\sim N$, the mutual purity becomes exponentially small in $N$; at this point, there is an $O(e^{-N})$ residual  correlation generated by the error which is unavoidable. 
\end{enumerate}
The third and final regime corresponds to an exchange of dominance between a leading saddle point and a subleading saddle point\footnote{Furthermore, there are also strictly subleading saddles controlled by a one-dimensional lattice of critical time points with the scrambling time as the lattice vector. The amount of correlation generated with the environment only changes as $T$ passes a lattice point.} in the Brownian SYK calculation,  analogously to the exchange of dominance between a disconnected geometry and the Euclidean wormhole in gravity. 
This quantitative hierarchy of complexity-protected error correction, ranging from a fragile encoding at $O(1)$ complexity, through a logarithmic complexity regime of reasonable protection, and finally an emergent robust error correction at large encoding complexity, is the central result of this paper. We regard this as a step towards understanding the structure of general entanglement wedges (Figure~\ref{fig:PL1}) from the boundary perspective in terms of quantum error correction.

Three sections follow. In Section~\ref{sec:prelim} we review the necessary ideas from quantum error correction. We discuss the class of errors of interest, and show that the mutual purity which is relevant for recovery from these errors can be expressed in terms of the standard purity\footnote{For a density matrix $\rho$, the purity is defined as $\mathrm{Tr}\,\rho^2$.} of a certain density matrix constructed using the encoding map.
We also briefly review the Brownian SYK model.
In Section~\ref{sec:typical-brownian}, we compute this  purity in the large $N$ limit using the Brownian SYK time evolution operator to model the encoding map.
We conclude with a discussion in Section~\ref{sec:disc}.
In Appendix~\ref{sec:Hamiltonian}, we give a Hamiltonian treatment of Brownian SYK to complement the path integral discussion in the main text and in Appendix~\ref{appQEC} we prove that the mutual purity provides a bound on the error correction properties of an encoding map.

\section{Setup}\label{sec:prelim}

\subsection{Brief review of quantum error correction}

The mathematical framework for quantum error correction involves an isometric embedding of a small ``code subspace'' $\cH_{\text{code}}$ into a larger Hilbert space $\cH_{\text{phys}}$: 
$$V:\cH_{\text{code}}\to \cH_{\text{phys}},$$
where $V^{\dagger}V = \mathbb{1}$. It is standard to model the error and recovery operations as completely positive trace-preserving linear maps, or ``quantum channels''.
Any such map $\cE$ has a representation in terms of its Kraus operators $\{E_m\}$ \cite{CHOI1975,NielsenChuang}:
\begin{equation}
    \cE(\rho) = \sum_m E_m \rho E_m^\dagger , \quad \sum_m E_m^\dagger E_m = \mathbb{1} .
\end{equation}
The minimum number of Kraus operators needed to implement a particular channel is called the rank of the channel.
These quantum channels act on physical density matrices, and the goal of error correction is to determine for a given error channel $\cE$ whether or not there exists a recovery channel $\cR$ which restores the state $\rho_\co$:
\begin{equation}
    \cR(\cE(V\rho_\co V^\dagger)) = \rho_\co .
\end{equation}
On the right hand side, we have in mind that the recovery channel has eliminated redundant portions of $\cH_\ph$,  leaving behind precisely the matrix $\rho_\co$ on the remaining subspace of $\cH_\ph$.

A second, convenient description of a quantum channel is given by its isometric extension, also known as its Stinespring dilation: we describe it as coupling the physical system via a unitary operator $U_{\cE}$ to  an auxiliary environment with Hilbert space $\cH_{\text{env}}$  spanned by basis elements $\{ |e_m\rangle_{{\rm env}} \}$, initially in some  fiducial state $\ket{e_0}$. The action of the channel $\cE$ on $\rho$ is then recovered by tracing out the environment: $\cE(\rho) = {\rm Tr}_{{\rm env}} \left[ U_{\cE} \,  ( \rho \otimes |e_0\rangle \langle e_0|_{{\rm env}} ) \, U_{\cE}^\dagger \right]$.  This implies that $E_m = \langle e_m | U_{\cE} | e_0\rangle$, or, equivalently,  
\begin{equation}
U_{\cE} |\psi\rangle \otimes |e_0\rangle = \sum_m E_m |\psi\rangle \otimes |e_m\rangle ,
\label{eq:UnitaryError}
\end{equation}
where $|\psi\rangle$ is any state in the physical Hilbert space. 

A standard fact in quantum error correction, sometimes called the {\it decoupling principle}, is that there always exists an approximate recovery channel where the error in recovery is bounded in terms of the amount of correlation the error channel generates between the code subspace and the environment. The convenient way to evaluate this correlation is to use the following procedure: (a) introduce a reference system $\cH_{{\rm ref}}$ which is isomorphic to and maximally entangled with with code Hilbert space, (b) act with the error quantum channel, (c) trace out the physical Hilbert space space, and (d) evaluate the correlation between the two remaining auxiliary spaces (the environment used to represent the channel and the reference space).  Thus, taking $\ket{i}_\re$ and $\ket{i}_\co$ to be orthonormal bases for the reference space and the code subspace respectively, we construct the state
\begin{equation}
    \ket{\Psi'} = \frac{1}{\sqrt{d_\co}}  \sum_{i=1}^{d_\co} \sum_{m=1}^{d_\en} \ket{i}_\re \otimes E_m V\ket{i}_\co \otimes \ket{e_m}_\en ,
    \label{eq:RefCode}
\end{equation}
where we have defined $d_X$ to be the dimension of a Hilbert space $\cH_X$.
Here the code states are embedded by $V$ into the physical Hilbert space and maximally entangled with the reference, while the error channel acts via $E_m$ on the physical states and thus entangles them with the environment.  Then we can say that  for any error channel $\cE = \{E_m\}$ there exists a recovery channel $\cR$ for which the Schatten 1-norm distance between the resulting state and the original encoded state is bounded as  \cite{Schumacher:1996dy,Beny2010}:
\begin{equation}
    \| \cR(\cE(V\rho_\co V^\dagger)) - \rho_\co \|_1 \leq \left( I_{\Psi'} (\re : \en) \right)^{1/4}  \, .
    \label{eq:MutualInfoBound}
\end{equation}
Here $I_{\Psi^\prime}({\rm ref}:{\rm env})$ is the mutual information between the environment and the reference space after tracing out the physical Hilbert space. This means that the error $\cE$ is exactly correctable in the code $V$ if the reference and environment do not share any correlation, hence the term ``decoupling''.

In this paper, we will be interested in quantum codes with complementary recovery \cite{Harlow:2016vwg}, which are the types of codes relevant for entanglement wedge reconstruction in AdS/CFT. For simplicity, consider a code subspace where we have some semi-classical bulk degrees of freedom in the entanglement wedge of a boundary subregion $A$, but no excitations in the entanglement wedge of the complement region $\overline{A}$. Let $\ket{i}_{\text{code}}$ denote basis states for these bulk degrees of freedom. It was shown by Harlow that the Ryu-Takayanagi formula together with quantum corrections implies the following structure for the encoding map in this situation:
\beq
 V: \cH_\co \to \cH_\ph , 
\eeq
\begin{equation} \label{harlow}
   V\ket{i}_\co = (U_A \otimes \mathbb{1}_\Abar ) \left( \ket{i}_{A_1} \otimes \ket{\chi}_{A_2\Abar} \right) ,
\end{equation}
where the physical Hilbert space (i.e., the Hilbert space of the dual CFT) is factorized as
\begin{equation}
    \cH_\ph = \cH_A \otimes \cH_{\Abar} , \quad \cH_A = \cH_{A_1} \otimes \cH_{A_2} \oplus \cH_{A_3} ,
    \label{eq:Adefs}
\end{equation}
and $\ket{\chi}$ is some fixed pure state in the Hilbert space $\cH_{A_2,\overline{A}}$. The argument for this involves the decoupling principle applied to the erasure of $\overline{A}$. Let us briefly recall how this works (see \cite{Harlow:2016vwg} for  details): we introduce an auxiliary system $\cH_{\text{aux}}$ isomorphic to the code subspace, and construct the state
\beq
|\Psi\rangle= \frac{1}{\sqrt{d_\co}}\sum_{i=1}^{d_\co} |i\rangle_{\text{aux}}\otimes |\psi_i\rangle_{A,\overline{A}},\;\;\;|\psi_i\rangle_{A,\overline{A}} = V|i\rangle_{\text{code}}.
\eeq
Since the bulk degrees of freedom in the code subspace are contained in the entanglement wedge of $A$, one can show using the RT \cite{Ryu:2006bv} plus FLM \cite{Faulkner:2013ana} formula that the mutual information $I(\text{aux} : \overline{A})$ vanishes, which implies that $\rho_{\text{aux},\overline{A}} = \rho_{\text{aux}}\otimes \rho_{\overline{A}}$. Therefore, viewed as a bipartite state on $A$ and $\text{aux} \, \cup \, \overline{A}$ the Schmidt vectors of $\Psi$ should take a factorized form on $\text{aux} \, \cup \, \overline{A}$. The \emph{canonical purification} \cite{Engelhardt:2018kcs, Dutta:2019gen} of $\rho_{\text{aux},\overline{A}}$ will therefore also have factorized states  on $A$. This is why the state inside the parentheses in equation \eqref{harlow} takes the factorized form between $A_1$ and $A_2$; here $A_1$ is the canonical purifier of $\text{aux}$ and $A_2$ is the canonical purifier of $\overline{A}$. Finally, any two purifications of the same density matrix $\rho_{\text{aux},\overline{A}} $ should be related by a unitary on $A$; this is precisely the unitary $U_A\otimes \mathbb{1}_{\overline{A}}$ appearing in equation \eqref{harlow}.
It is easy to check that this code subspace is protected against the erasure of $\overline{A}$.
We will refer to the operator $U_A$ as the \textit{encoding unitary}.\footnote{The additional Hilbert space component $\cH_{A_3}$ in \eqref{eq:Adefs} is, for our purposes, a bookkeeping device for situations where the physical Hilbert space dimension is not a product of integers, also implying that a part of it does not participate in the code; so we will simply drop $\cH_{A_3}$ as it is not pertinent to our considerations. Henceforth, we will focus our attention on codes which have the above structure, but without $\cH_{A_3}$.}

There is an important caveat: the bulk-to-boundary map $V$ need not be an exact isometry, and is often an approximate one with corrections of $O(e^{-1/G_N})$.\footnote{There are also more extreme situations in which the map is far from an isometry \cite{Akers:2022qdl,Kar:2022qkf,Balasubramanian:2022fiy}.}
Relatedly, the quantum generalization of the Ryu-Takayanagi formula, namely the QES formula, is correct to all orders in the $G_N$ perturbation theory for appropriate states,\footnote{See \cite{Akers:2020pmf} for situations where there are leading order corrections.} but in general there are corrections of $O(e^{-1/G_N})$. Therefore, the bulk-to-boundary map in AdS/CFT is only approximately of the form \eqref{harlow}, and has additional exponentially small corrections. As a first pass, we will focus on codes of the type \eqref{harlow} in this work. It would be interesting to incorporate the corrections mentioned above in our analysis, but we will not attempt this here.

\subsection{An error correction bound}
Putting together the considerations from above, we first introduce a reference system isomorphic to the code subspace and consider the maximally entangled state:
\beq
|\Psi\rangle = \frac{1}{\sqrt{d_\co}}\sum_i |i\rangle_{\text{ref}}\otimes U_A \left(|i\rangle_{A_1}\otimes |\chi\rangle_{A_2\Abar}\right)\otimes |e_0\rangle_{\en},
\label{eq:InitialState}
\eeq
where we included the code subspace structure in \eqref{harlow} and an auxiliary environment in some fiducial initial state $\ket{e_0}$.  The error now acts in the form of a joint unitary operator on $A \, \cup \, \text{env}$ (we assume the error does not act on the $\Abar$ system):
\beq
|\Psi'\rangle = \frac{1}{\sqrt{d_\co}} \sum_i |i\rangle_{\re}\otimes U_{\cE}\left[ U_A \left(|i\rangle_{A_1}\otimes |\chi\rangle_{A_2\Abar}\right)\otimes |e_0\rangle_{\en}\right] ,
\eeq
where we applied the error channel as in \eqref{eq:RefCode} to the state \eqref{eq:InitialState} in terms of a unitary operator \eqref{eq:UnitaryError} entangling the physical system with the environment.  Next, we  obtain the reduced density on the reference and environment subsystems:
\beq
\rho'_{\re,\en}= \frac{1}{d_\co} \sum_{i,j}|i\rangle\langle j|_{\re}\otimes \mathrm{Tr}_A\left\{U_{\cE}\left[U_A\left(|i\rangle\langle j|_{A_1}\otimes \rho^{\chi}_{A_2}\right)U_A^{\dagger}\otimes |e_0\rangle\langle e_0|_{\en}\right]U^{\dagger}_{\cE}\right\},
\label{eq:reduced-ref-env-encoding}
\eeq
where  we have performed the trace over $\overline{A}$ and replaced $\chi$ with its reduced density matrix $\rho^{\chi}_{A_2}$.  Finally, following \eqref{eq:MutualInfoBound} we can bound the error in recovery of the original state after action of the error channel in terms of the fourth root of the mutual information between the reference and the environment:
\begin{equation}
    I_{\Psi'}(\re : \en) = S(\rho'_{\re}) + S(\rho'_{\en}) - S(\rho'_{\re,\en}) ,
    \label{eq:psiprimemutualinfo}
\end{equation}
where the von Neumann entropies on the right hand side are computed from $\rho'_{\re,\en}$ and the reduced density matrices on the reference and the environment $\rho'_{\re} = {\rm Tr}_\en(\rho'_{\re,\en})$ and $\rho'_{\en} = {\rm Tr}_\re(\rho'_{\re,\en})$.

The mutual information in \eqref{eq:psiprimemutualinfo} is difficult to compute directly. A standard approach is to use the replica trick to obtain the von Neumann entropies on the right hand side as analytic continuations of the R\'{e}nyi entropies which are easier to compute via the relation
\begin{equation} 
S(\rho) = -{\rm Tr}(\rho \log \rho) = \lim_{n\to1} \frac{1}{1-n} S^{(n)}(\rho) ,
\end{equation} 
where $S^{(n)}(\rho) = \log {\rm Tr}(\rho^n)$ is the $n^{{\rm th}}$ R\'{e}nyi entropy.
We will take a different approach. In Appendix~\ref{appQEC} we study a particular, well-motivated recovery channel and show that the trace distance between the recovered state under this recovery channel and the actual state satisfies 
\begin{equation}\label{eq:Renyi2Bound}
D(\rerr{V\rho_\co V^\dagger},\rho_\co) 
    \leq c\left( \Tr\left( \rho'^2_{\re,\en} - \rho'^2_{\re} \otimes \rho'^2_{\en} \right)\right)^{1/4} ,\;\;\;\;c= d_\co^{5/2}\, d_\en^{1/2},
\end{equation}
where $D(\rho,\sigma) = \frac{1}{2} \Tr( | \rho - \sigma | ) $ with $|X| = \sqrt{X^\dagger X}$ is the trace distance between density matrices. As above, $d_\co$ and $d_{{\rm env}}$ are dimensions of the code/reference subspace and the environment in the isometric extension of the error channel, respectively. 
In this work, \eqref{eq:Renyi2Bound} will replace the standard decoupling principle \eqref{eq:MutualInfoBound} due to the ease of evaluating the right hand side.
In particular, the expression \eqref{eq:Renyi2Bound} bounds the error in recovery directly in terms of the quantity
\begin{equation}
\label{MIexp}
\mathcal{F}_{\Psi'}(\re:\en) \equiv \Tr\left( \rho'^2_{\re,\en} - \rho'^2_{\re} \otimes \rho'^2_{\en} \right) ,
\end{equation}
which we call the \textit{mutual purity}.
If $\mathcal{F}$  vanishes, so does the right hand side of the bound  \eqref{eq:Renyi2Bound}, so that perfect recovery is possible and the error can be corrected. In view of this bound, below we will compute $\mathcal{F}$ to quantify the robustness against errors for encoding maps of increasing complexity.

\subsection{Error correction and maximum complexity encoding}

To get a more quantitative understanding of what happens when the encoding unitary becomes complex, as a first pass we can compute the Haar ensemble average  with respect to $U_A$ of $\mathcal{F}_{\Psi'}$. This is because we expect that the typical unitary in the Haar ensemble will be exponentially complex, and that as long as the dimension of $\mathcal{H}_A$ is large, deviations away from the ensemble average will be exponentially suppressed in the number of qubits. 
For a Haar random unitary $U$ acting on a Hilbert space $\cH_X$, a standard formula for Haar integration says:
\beq\label{eq:Haaraverage}
\langle U_{m_1p_1} U^{\dagger}_{q_1n_1}U_{m_2p_2} U^{\dagger}_{q_2n_2}\rangle_{\text{Haar}} =\frac{1}{d_X^2}\left(\delta_{m_1,n_1}\delta_{m_2,n_2}\delta_{p_1,q_1}\delta_{p_2,q_2}+\delta_{m_1,n_2}\delta_{m_2,n_1}\delta_{p_1,q_2}\delta_{p_2,q_1}\right)+O(\frac{1}{d_X^3}) .
\eeq
This expression has a gravitational analogue: in the Euclidean path integral computation of the radiation purity in the PSSY toy model for an evaporating black hole in JT gravity \cite{Penington:2019kki}, the two terms displayed above are respectively analogous to the ``disconnected'' and ``wormhole'' gravitational saddles. Using the above integral we can now evaluate the Haar average of $\mathrm{Tr}_{\re,\en}\left[(\rho'_{\re,\en})^2\right]$ as follows.
\begin{equation}
\begin{split}
 \avg{ \Tr \rho'^2_{\re,\en}}_{\text{Haar}} &= \frac{1}{d_\co^2} \sum_{i,j,m,n} \avg{ \mathrm{Tr}_A\left\{E_m \,U_A\left(|i\rangle\langle j|_{A_1}\otimes \rho^{\chi}_{A_2}\right)U_A^{\dagger} \, E_n^{\dagger} \right\} 
\\ 
&\hspace{3cm}\mathrm{Tr}_A\left\{E_n \, U_A\left(|j\rangle\langle i|_{A_1}\otimes \rho^{\chi}_{A_2}\right)U_A^{\dagger}\,E_m^\dagger \right\} }_{\text{Haar}} \\  &\approx \frac{1}{d_\co^2\, d_A^2}\sum_{i,j,m,n}  \Tr ( E_m^\dagger E_n )\, \delta_{i,j} \Tr(\rho_{A_2}^{\chi}) \Tr (E_n^\dagger E_m)\, \delta_{j,i} \Tr(\rho_{A_2}^\chi) \\ &\hspace{3cm}  +  \frac{1}{d_\co^2\, d_A^2}\sum_{i,j,m,n} \Tr(E_m E_m^\dagger E_n E_n^\dagger) \Tr(\rho^{\chi\, 2}_{A_2}) \,\delta_{i,i} \,\delta_{j,j}
   \\ &= \frac{1}{d_\co^2\, d_A^2}\sum_{m,n}  \left(d_\co \Tr ( E_m^\dagger E_n ) \Tr (E_n^\dagger E_m)  +  d_\co^2 \Tr(E_m E_m^\dagger E_n E_n^\dagger) \Tr(\rho^{\chi\, 2}_{A_2}) \right)  \\ 
   &= \frac{1}{d_\co^2 d_A^2} \left( d_\co\, d_A^2 \Tr (\sigma_\en^2) + d_\co^2 \, d_A^2\Tr(\sigma_A^2)\Tr(\rho^{\chi\, 2}_{A_2}) \right) \\
   &= \frac{ \Tr (\sigma_\en^2)}{ d_\co} \left( 1 + d_\co \frac{\Tr (\sigma_A^2) \Tr(\rho^{\chi\, 2}_{A_2})}{\Tr (\sigma_\en^2)} \right) ,
\end{split} 
\end{equation}
where we have defined the density matrix $\sigma$ on $\cH_A \otimes \cH_\en$ as
\beq
\sigma \equiv 
U_{\cE} \left( \frac{\mathbb{1}_A}{d_A}\otimes |e_0\rangle\langle e_0|_{\en} \right) U^{\dagger}_{\cE} 
= 
\sum_{m,n} E_m \frac{\mathbb{1}_A}{d_A}E_n^\dagger \otimes \ket{e_m} \bra{e_n} ,
\label{eq:sigma-A-env}
\eeq
and the associated reduced density matrices
\begin{equation}\label{eq:sig_a_env}
\sigma_A = \mathrm{Tr}_{\en}\,\sigma = \sum_{m} E_m \frac{\mathbb{1}_A}{d_A}E_m^\dagger\ ,\quad 
\sigma_{\en}=\,\mathrm{Tr}_A\,\sigma \; = \sum_{m,n} \frac{\Tr_A \left(E_m E_n^\dagger\right)}{d_A} \ket{e_m} \bra{e_n} \, .
\end{equation}
In the first step, we represented the action of $U_\mathcal{E}$ in terms of the Kraus operators $E_m = \langle e_m | \,U_{\cE} | e_0\rangle$ and traced out the reference and the environment degrees of freedom. The second step follows from equation (\ref{eq:Haaraverage}). In the fourth step, we used equations (\ref{eq:sigma-A-env}) and (\ref{eq:sig_a_env}). 
Following similar steps, we can compute the Haar average of  $\Tr \rho'^2_{\en}$:
\begin{equation}
 \avg{ \Tr \rho'^2_{\en}}_{\text{Haar}} 
   \approx \Tr (\sigma_\en^2) \left( 1 +  \frac{\Tr (\sigma_A^2) \Tr(\rho^{\chi\, 2}_{A_2})}{d_\co \, \Tr (\sigma_\en^2)} \right) .
\end{equation}
Since $\rho'_\re$ is maximally mixed, we have
\begin{equation} 
\Tr \rho'^2_\re = \frac{1}{d_\co} ,
\end{equation}
for any $U_A$, which means this expression factors out of any Haar average since it is independent of $U_A$.
Combining the above results for $\avg{ \Tr \rho'^2_{\re,\en}}_{\text{Haar}}$, $\avg{ \Tr \rho'^2_{\en}}_{\text{Haar}} $, and $\Tr \rho'^2_\re$, the final result for the Haar averaged mutual purity is given by 
\begin{equation}
    \avg{\cF_{\Psi'}(\re : \en)}_{\text{Haar}} =   e^{-S^{(2)}(\sigma_A) - S^{(2)}(\chi_\Abar)}\left(1 - \frac{1}{d_\co^2}\right)  + \cdots ,
\label{eq:nth-renyi-mutual2}
\end{equation}
where the $\cdots$ indicate  exponentially small contributions that we have dropped along the way, 
$S^{(2)}(\sigma_A)$ is the second R\'enyi entropy of the $A$ subsystem in the mixed state $\sigma$, and $S^{(2)}(\chi_\Abar)$ is the second R\'enyi entropy of the $\overline{A}$ subsystem in the state $\ket{\chi}_{A_2 \overline{A}}$.
The salient feature of \eqref{eq:nth-renyi-mutual2} is the leading exponential suppression, as we will now describe.
The quantity $1 - d^{-2}_\co$ is simply an $O(1)$ prefactor for a nontrivial code subspace.

Two features of  \eqref{eq:nth-renyi-mutual2}  are worth highlighting. Firstly, note from the final formula that in the typical code drawn from the Haar ensemble, the error channel perceives the state on $A$ as maximally mixed, and gains no access to the microscopic structure of the state. 
Consequently, as long as the error channel is low-rank, we see that $\avg{\cF_{\Psi'}(\re : \en)}_{\text{Haar}}$ is exponentially suppressed by $e^{-S^{(2)}(\sigma_A)}$.  This is a direct consequence of complexity -- a general, complex encoding unitary scrambles the code subspace to a point where generic error channels do not gain any access to it. (A similar coarse-graining picture for apparent horizons and quantum extremal surfaces was advocated in \cite{Engelhardt:2018kcs, Engelhardt:2021mue, Chandra:2022fwi}.) Furthermore, there is an additional suppression factor of $e^{-S^{(2)}(\chi_\Abar)}$ in equation \eqref{eq:nth-renyi-mutual2} coming from the shared entanglement with $\overline{A}$. The combination of these two effects coming from complexity and entanglement thus makes the code robust against generic, low-rank errors.

In the above analysis, we have assumed that the error channel does not have prior access to the encoding unitary $U_A$. This is crucial, because with prior access to the details of the encoding unitary, it is possible to construct low-rank error channels which corrupt the code subspace. For example, consider the error channel:
\beq
\cE(\rho_A) = \cE_{\text{partial SWAP}}(U_A^{\dagger}\rho_A U_A),
\eeq
where the unitaries $U_A^{\dagger} (\cdots) U_A$ first undo the encoding, and the partial swap then swaps out the state on the first $t$ qubits with the environment. Since the reference system in $\Psi'$ is maximally entangled with the qubits in $A_1$, even if the partial SWAP acts on one of the qubits in $A_1$, then it will generate an $O(1)$ amount of mutual information between $\cH_\re$ and $\cH_\en$, and thus error correction fails. There is an analogue of this in the JT gravity model \cite{Balasubramanian:2022fiy} --- there, one assumes that the error channel does not ``generate'' additional asymptotic boundaries which can connect up with the bulk geometry and modify the mutual information. Of course, note that this channel is fine-tuned, in that it uses the specific unitary $U_A$ which goes into the encoding. Nevertheless, if $U_A$ is computationally simple, then the above error channel is also simple. On the other hand, when the encoding unitary $U_A$ is exponentially complex, the error channel described above must be equally complex in order to first undo the encoding. Thus, if the python's lunch has an exponentially complex encoding map, then although it will not not be robust against the error channels which are constructed with prior access to the encoding unitary, the channel in question will be exponentially complex. So it will be extremely difficult to implement such errors. This is again a manifestation of the idea that semi-classical causality in the bulk is robust due to complexity.

\subsection{Random circuit codes: Brownian SYK}

Our goal in the rest of the paper is to study in more detail the dependence of the error correction against generic, low-rank errors acting on $A$ relative to the complexity of the encoding unitary. It is convenient, for this purpose, to study the ensemble average of the mutual purity introduced above, but we would like to consider a one-parameter family of ensembles, labelled by the complexity of the typical unitary in the ensemble. 

A simple way to generate such an ensemble is to consider the time evolution operators $U_A = e^{-iTH}$, for some ensemble of chaotic Hamiltonians. It is important that the Hamiltonians be chaotic, because for integrable Hamiltonians, the complexity of the time evolution operator is expected to saturate at a sub-exponential time-scale \cite{Balasubramanian:2021mxo}.  On the other hand, for chaotic Hamiltonians, it is widely expected that the complexity $\mathcal{C}(e^{-iTH})$ grows linearly with time $T$ for an exponential amount of time: $\mathcal{C}(e^{-iTH}) \propto T$ (examples in \cite{Balasubramanian:2019wgd,Balasubramanian:2021mxo,Balasubramanian:2022tpr,Balasubramanian:2022dnj}).  Thus, the parameter $T$ is expected to be a good measure of the complexity for chaotic Hamiltonians for exponentially long times.
Considering an ensemble of chaotic Hamiltonians then allows us to rely on this property holding only for the typical chaotic Hamiltonian, which is a much weaker assumption than expecting an arbitrary chaotic Hamiltonian to have linearly growing complexity.

More generally, we could consider unitaries $U_A$ which are constructed from random circuits. Any unitary can be constructed as a circuit with local quantum gates --- in a random circuit, we randomly choose the local gates at each instant of time from some ensemble. The resulting one-parameter family of random circuit ensembles may be thought of as a one-parameter family of measures $\d\mu (T)$ on the unitary group $U(d_A)$.
To guarantee increasing complexity, we can choose $\d\mu (T)$ to be highly concentrated at the identity when $T = 0$, and as $T$ increases we require that the support of $\d\mu (T)$ should expand outward on $U(d_A)$ like a gas, eventually covering the entire group.
If we further require that $\d\mu (T)$ approaches the Haar measure when $T \to \infty$, we can guarantee that the typical operator selected by averaging with $\d\mu (T)$ will have roughly increasing complexity as $T$ increases \cite{Jian:2022pvj}.

To define such a one-parameter ensemble of random circuit codes it is convenient to pick $U_A(T)$ to be the time evolution operator of the Brownian SYK model \cite{Kitaev2015v1,Kitaev2015v2,Maldacena:2016hyu,Saad:2018bqo,Stanford:2021bhl}.
This model is constructed from $N$ Majorana fermions $\psi_a$, and is defined by a set of random couplings:
\begin{equation}
    H(t) = i^{q/2} \sum_{1\leq a_1<\dots <a_q\leq N} J_{a_1\dots a_q}(t) \psi_{a_1} \dots \psi_{a_q} , \quad \{\psi_a,\psi_b\} = \delta_{ab} .
\end{equation}
The coupling constants are time-dependent and are chosen to be independently Gaussian at each time point with mean zero and a fixed variance:
\begin{equation}
    \avg{J_{a_1\dots a_q}(t) J_{b_1\dots b_q}(t') } = \delta_{a_1b_1} \dots \delta_{a_qb_q} \frac{(q-1)!}{N^{q-1}} J^2(t,t') , \quad J^2(t,t') = J\delta(t-t') .
\end{equation}
The associated encoding unitary operator is
\begin{equation}
    U_A(T) = \mathcal{T} \exp \left( -i \int_0^T \d t\; H(t) \right) ,
\label{eq:brownian-code}
\end{equation}
where $\mathcal{T}$ is the time-ordering operator.
Note that we are using the Brownian theory not as a model of a holographic boundary theory Hamiltonian (as has been done previously \cite{Stanford:2021bhl}), but rather as the generator of a family of holographic encoding (bulk-to-boundary) maps.
Because $H(t)$ depends on random couplings, $U_A(T)$ is a random variable which has support on certain portions of the unitary group depending on the magnitude of $T$.
The relevant portions are analogous to regions of space covered by a random walk of a certain fixed length.

An subtlety which we will return to later is that the SYK theory obeys certain global symmetries.
The presence of these symmetries prevents the effective measure $\d\mu (T)$ from covering the entire unitary group as $T\to \infty$.
To get around this, we will follow the strategy of \cite{Stanford:2021bhl}, where a semi-classical analysis of the SYK theory gave a natural way of extracting results for SYK-like theories which do end up covering the whole unitary group.

\subsection{Erasure errors}\label{ssec:typical-errors}

In order to further simplify the problem, we will consider a particular class of errors. It is important that the error channel has no prior access to the encoding unitary, i.e., we want the error to be generic and low-rank. The error channel we  consider in this work will be the erasure of some subsystem $R$.

Let us define
\begin{equation} 
\sigma \equiv \frac{1}{d_\co} VV^\dagger = \frac{1}{d_\co}\sum_i |\psi_i\rangle\langle \psi_i|, 
\end{equation}
and let $\sigma_R$ be the reduced density matrix on $R$, and $\sigma_L$ be the corresponding reduced density matrix on the rest of the system $L$. 
We have defined this new $\sigma$, which we will use for the rest of this paper, in place of the previous one in \eqref{eq:sigma-A-env}.
Then, for such an erasure error, 
\beqn\label{MItyp'}
\mathcal{F}_{\Psi'}(\re:\en) &=& \mathcal{F}_{\Psi'}(\re:R)\nonumber\\
&=&\Tr\sigma_L^2-\frac{1}{d_\co}\Tr\,\sigma_R^2 ,
\eeqn
where recall that $\ket{\Psi'}$ was the state which resulted from applying the error to the maximally entangled state between the reference and the code subspace. For simplicity, we will specialize to the case where $R=A_1$ and $L=A_2\cup \Abar$. In addition, using the fact that $|\chi\rangle_{A_2\Abar}$ is maximally entangled, we then arrive at
\beq
\mathcal{F}_{\Psi'}(\re:R)=\left(\Tr\sigma_L^2-\frac{1}{d^2_\co}\right).
\eeq
Since the dimension of the environment in this case is the same as $d_{\co}$, we find that the robustness of error correction for the above erasure error is bounded by the quantity $d_{\co}\left(d_{\co}^2\Tr\sigma_L^2-1\right)$.

\begin{figure}[!t]
    \centering
    \includegraphics[scale=.75]{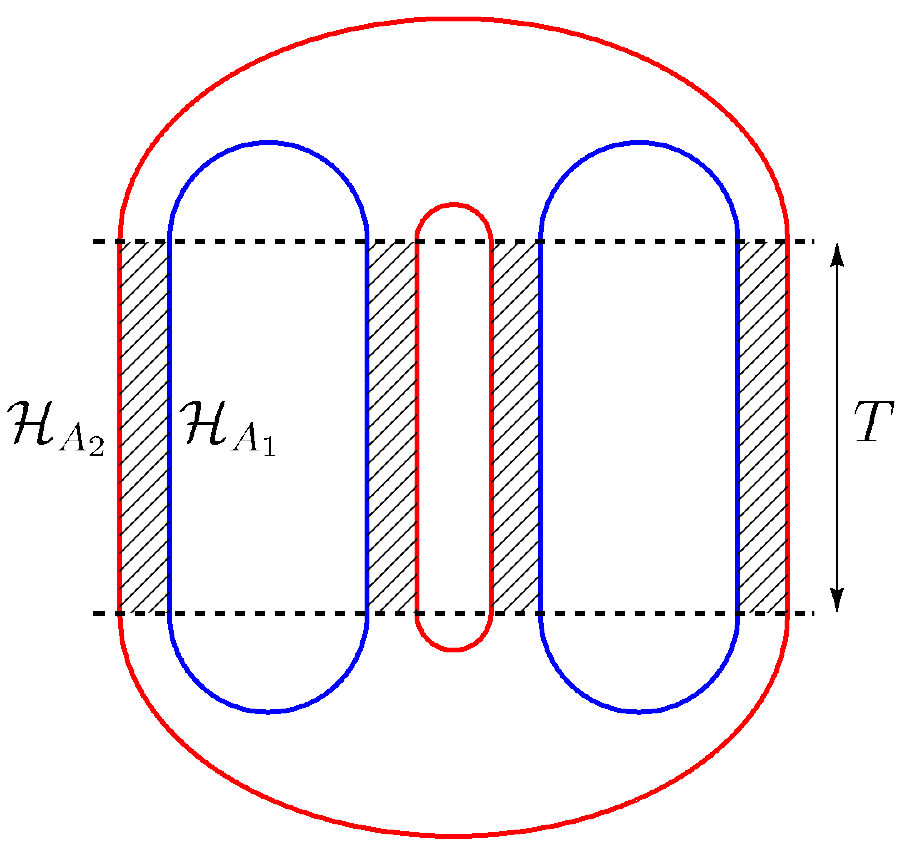}
    \caption{\small{The contour computing $\Tr \sigma_L^2$, the quantity relevant for the mutual purity for the erasure of $A_1$. The red contour corresponds to the $A_2$ fermions while the blue corresponds to the $A_1$ fermions. The hatched regions denote an application of the time evolution operator $U_A(T)$ or $U_A^\dagger(T)$ of Brownian SYK, which couples the $A_1$ and $A_2$ systems.  We have omitted the contour orientations which determine the forward and backward time evolution, but from left to right the hatched regions alternate between $U_A^\dagger(T)$ and $U_A(T)$, beginning with $U_A^\dagger(T)$. The arcs at the top and the bottom specify the final and initial conditions respectively; in our calculation, all these arcs are actually infinitesimally small (corresponding to maximal entanglement), but they have been enlarged for visual clarity.}}
    \label{fig:contour2}
\end{figure}

In the next section, we will turn to the main objective of this work: computing the purity $\Tr\sigma_L^2$ for Brownian SYK codes. In particular, we are interested in the dependence of this quantity on the encoding complexity of the code, which as explained above, is linearly related to the time parameter $T$. From equation \eqref{MItyp'},  we need to compute  $\Tr\sigma_L^2$ as a function of $T$. Actually, since $\Abar$ has no dynamics associated with it (i.e., there is no non-trivial time evolution operator acting on $\Abar$), this computation boils down to a Lorentzian path integral entirely in the $A_1A_2$ subsystem --- the relevant time contours are shown in Figure~\ref{fig:contour2}. 
To arrive at Figure~\ref{fig:contour2}, we notice that $\Tr \sigma_L^2$ involves two copies of $U_A$ and two copies of $U_A^\dagger$, and so can be thought of as a matrix element of the operator $U_A^\dagger \otimes U_A \otimes U_A^\dagger \otimes U_A$.
The matrix element in question is determined by the trace structure: since the $R = A_1$ system is traced out first to obtain $\sigma_L$, the adjacent blue $A_1$ contours are joined in Figure~\ref{fig:contour2}, while the secondary trace over $L$ joins the inner and outer red $A_2$ contours.

When $T$ is small, we expect $\mathrm{Tr}\,\sigma_L^2$ to be close to 1, and so the mutual purity is non-zero. On the other hand, at very late times, we expect $\mathrm{Tr}\,\sigma_L^2$ to approach $1/d_\co^2$ and the mutual purity to approach zero. The intuitive argument for this is as follows: let us first purify the density matrix $\sigma$ by including an auxiliary system $\text{aux}$ which is isomorphic to the code subspace:
\beqn \label{eq:expec}
|\psi_{\sigma}\rangle &=& \frac{1}{\sqrt{d_\co}}\sum_i |i\rangle_{\text{aux}}\otimes |\psi_i\rangle \nonumber\\
&=&\frac{1}{\sqrt{d_\co}}\sum_i |i\rangle_{\text{aux}}\otimes U_{A}(T)|i\rangle_{A_1}\otimes |\chi\rangle_{A_2,\Abar}.
\eeqn
When $T=0$, the subsystem $A_1$ is maximally entangled with $\text{aux}$ while $L=A_2\cup \Abar$ is in a pure state. When $T$ becomes large (on the order of the scrambling time), we expect $U_A(T)$ to generate nearly maximal entanglement between $A_1$ and $L$. By the monogamy of entanglement, therefore, $A_1$ cannot share much entanglement with $\text{aux}$. However, the unitary operator never acted on $\text{aux}$; thus the reduced density matrix on $\text{aux}$ must still be maximally mixed. We therefore conclude that both $A_1$ and $\text{aux}$ are close to being in a maximally entangled state with $L$, and so the purity of $L$ must approach $\frac{1}{d_\co^2}$. Consequently, $\mathcal{F}_{\Psi'}(\re:R)$ should approach zero.  

In what follows, we wish to understand the detailed time-dependence of the mutual purity at late times. In particular, we will demonstrate that the mutual purity becomes $O(\frac{1}{N})$ by the scrambling time $T\sim \frac{1}{J} \log N$, but then continues to decay thereafter, approaching its saturation value which is $O(e^{-N})$ at a time of order $T\sim \frac{1}{J}N$. The important point is that the mutual purity keeps decaying even beyond the scrambling time, until it reaches an exponentially small plateau, which in the present model happens at polynomial time.\footnote{It is plausible that the time-scale at which the saturation happens is an artefact of the ensemble we have chosen, and that for other choices of ensembles, the plateau happens at exponential times.} We will interpret this phenomenon as ``complexity-protected quantum error correction''.

\section{Erasures in Brownian SYK codes}\label{sec:typical-brownian}

\subsection{Boundary conditions and large $N$ equations}\label{ssec:boundary-conditions-brownian}

Our task now is to evaluate the path integral of Brownian SYK on the contour in Figure~\ref{fig:contour2}.
Following \cite{Stanford:2021bhl}, we will use the collective-variable description of Brownian SYK. We view the path integral in Figure~\ref{fig:contour2} as an amplitude where we start with an ``in'' state, then time evolve for a time $T$ and then take the overlap with an ``out'' state. The boundary  conditions relevant for us are  as follows. For the in boundary conditions, we have
\beq
\psi^{(1)}_{a_1}|\text{in}\rangle = i\psi^{(2)}_{a_1}|\text{in}\rangle,\;\;\psi^{(3)}_{a_1}|\text{in}\rangle = i\psi^{(4)}_{a_1}|\text{in}\rangle ,
\eeq
\beq
\psi^{(1)}_{a_2}|\text{in}\rangle = i\psi^{(4)}_{a_2}|\text{in}\rangle,\;\;\psi^{(2)}_{a_2}|\text{in}\rangle = i\psi^{(3)}_{a_2}|\text{in}\rangle,
\eeq
while for the out state we have the adjoint boundary conditions:
\beq
\langle \text{out}|\psi^{(1)}_{a_1} = -i\langle \text{out}|\psi^{(2)}_{a_1},\;\;\langle \text{out}|\psi^{(3)}_{a_1} = -i\langle \text{out}|\psi^{(4)}_{a_1} ,
\eeq
\beq
\langle \text{out}|\psi^{(1)}_{a_2} = -i\langle \text{out}|\psi^{(4)}_{a_2},\;\;\langle \text{out}|\psi^{(2)}_{a_2} = -i\langle \text{out}|\psi^{(3)}_{a_2}.
\eeq
Here $a_1$ denotes the index of the $N_1$ fermions corresponding to the subsystem $A_1$, while $a_2$ denotes the  index of the $N_2$ fermions corresponding to the subsystem $A_2$. The superscript index $(i)$ on $\psi^{(i)}_a$ (where $i=1,\cdots,4$) labels the four contour segments corresponding to real time evolution.
From left to right in Figure~\ref{fig:contour2}, we label the contours 1, 2, 3, and 4.

In order to evaluate the path integral, it is convenient to define the two matrices:
\beq
g^{(1)}_{ij}(t) = \frac{1}{N_1}\sum_{a_1} \langle \psi_{a_1}^{(i)}(t)\psi^{(j)}_{a_1}(t)\rangle,\;\;g^{(2)}_{ij}(t) = \frac{1}{N_2}\sum_{a_2} \langle \psi_{a_2}^{(i)}(t)\psi^{(j)}_{a_2}(t)\rangle,
\eeq
which we can think of as the singlet part of the fermion two-point functions in the $A_1$ and  $A_2$ sectors respectively.  Here we inserted the operators on the right hand side at the specified time into the path integral in Figure~\ref{fig:contour2}.
We will soon see that these two sets of variables control the classical limit of the Brownian theory on this contour. To solve the classical equations of motion we will obtain in this limit,  we require the boundary conditions that are implied by the in and out state relations above.
It is also convenient to define the total two-point function (i.e., the summed two-point function of all the fermions):
\beq
g_{ij} = \lambda g^{(1)}_{ij} + (1-\lambda)g^{(2)}_{ij},
\eeq
where we have introduced the parameter $\lambda = \frac{N_1}{N}$. When evaluating the path integral at large $N$, it will be convenient to take the  double scaling limit:
\begin{equation} 
N_1 \to \infty,\;\; N_2 \to \infty,\;\; \lambda = \frac{N_1}{N}\;\text{fixed}.
\label{eq:classical-limit}
\end{equation}
In fact, $A_1$ has the same dimension as the code subspace, so we would like to take $N_1$ much smaller than $N_2$. This corresponds to taking $\lambda \ll 1$. Thus, we can take $\lambda$ to be a small (but $O(N^0)$) parameter and work in perturbation theory in $\lambda$. This makes some of the path integral calculations analytically tractable. 

Note that both the in and out boundary conditions have the  property that (recall that the fermions are normalized such that $\psi^2 = 1/2$):
\beq \label{parity1}
\psi^{(1)}_{a_1}\psi^{(2)}_{a_1}\psi^{(3)}_{a_1}\psi^{(4)}_{a_1} |\text{in}\rangle = -\frac{1}{4} \ket{\text{in}},
\eeq
\beq \label{parity2}
\psi^{(1)}_{a_2}\psi^{(2)}_{a_2}\psi^{(3)}_{a_2}\psi^{(4)}_{a_2} |\text{in}\rangle = - \frac{1}{4} \ket{\text{in}}.
\eeq
Since $A_1$ fermions lie in the same parity sector as the $A_2$ fermions, and the (effective) Hamiltonian commutes with the fermion parity operator after averaging (see Appendix \ref{sec:Hamiltonian}, equation \eqref{eq:typhamiltonian} and discussion below it),  the above relations should hold at any time. Equations \eqref{parity1} and \eqref{parity2} imply the following symmetry properties:
\beq
g_{12}^{(1)} = g_{34}^{(1)},\;\; g^{(1)}_{14} = g^{(1)}_{23},\;\; g^{(1)}_{24} = - g^{(1)}_{13},
\eeq
\beq
g_{12}^{(2)} =  g_{34}^{(2)},\;\; g^{(2)}_{14} = g^{(2)}_{23},\;\; g^{(2)}_{24} =  -g^{(2)}_{13}.
\eeq
These should hold at all times because time evolution preserves fermion parity flavor-wise. So the evolution reduces to the six variables $x_{\alpha} = 2ig^{(\alpha)}_{12}$, $y_{\alpha} = 2g^{(\alpha)}_{13}$ and $z_{\alpha} = 2ig^{(\alpha)}_{14}$, where $\alpha=1,2$. We can now rewrite the initial and final boundary conditions in terms of these new variables as
\beq
x_1(0)= 1,\;\;y_1(0) = z_1(0),
\label{eq:typbc1}
\eeq
\beq
x_2(0)= -y_2(0),\;\;z_2(0)=1.
\label{eq:typbc2}
	\eeq
	\beq
	x_1(T)= 1,\;\;y_1(T) = - z_1(T),
	\label{eq:typbc3}
\eeq
\beq
x_2(T)= y_2(T),\;\;z_2(T)=1.
\label{eq:typbc4}
\eeq
It is convenient to also introduce the total variables $x =\lambda x_1+ (1-\lambda)x_2$, and similarly for $y$ and $z$. The above boundary conditions imply the following constraints in terms of the $(x,y,z)$ variables:
\beq
z(0)-x(0)-y(0)= (1-2\lambda),\;\;\;z(T)-x(T)+y(T)= (1-2\lambda).
\eeq

In order to proceed with the evaluation of the Lorentzian path integral in Figure~\ref{fig:contour2}, recall \cite{Stanford:2021bhl} that the action for the Brownian SYK model on the contour in Figure~\ref{fig:contour2} is given by
\beq
I = \frac{1}{2} \int_0^T \d t \,\left( \psi^{(j)}_a \,  \partial_t  \psi^{(j)}_a  +i^{\frac{q}{2}} \, s_j \, J_{a_1\cdots a_q} \, \psi^{(j)}_{a_1\cdots a_q}\right),
\eeq
where the flavor indices run over $a = 1, \cdots, N$ (i.e., over both $A_1$ as well as $A_2$ fermions), and we have introduced the notation $\psi^{(j)}_{a_1\cdots a_q} = \psi^{(j)}_{a_1}\cdots \psi^{(j)}_{a_q}$. 
The quantity $s_j$ is given by
\begin{equation}
    s_j = \begin{cases}
    +i , & j \in \{2,4\} \\
    -i \cdot i^q , & j \in \{1,3\} ,
    \end{cases}
\end{equation}
and is related to the difference between forward and backward time evolution (see \cite{Stanford:2021bhl} for details).
We now wish to perform the average over the couplings. Using 
\beq
\langle J_{a_1\cdots a_q}(t) J_{b_1\cdots b_q}(t')\rangle = \delta_{a_1b_1}\cdots \delta_{a_qb_q} \frac{(q-1)!}{N^{q-1}} J^2(t,t'),
\eeq
the action obtained after ensemble averaging over the couplings is given by\footnote{In the second step, we have made the same imprecise replacement of the Hamiltonian  as in \cite{Stanford:2021bhl}, discussed in more detail in Appendix A.3 of \cite{Saad:2018bqo}.}
	\beqn
	I &=&\frac{1}{2} \int_0^T \d t \, \psi^{(j)}_a \partial_t  \psi^{(j)}_a - \frac{i^q(q-1)!}{2N^{q-1}} \iint_0^T \d t \d t' J^2(t,t') s_j s_{j'} \,\psi^{(j)}_{a_1\cdots a_q}(t) \psi^{(j')}_{a_1\cdots a_q}(t') \nonumber\\
	&=&\frac{1}{2} \int_0^T \d t \, \psi^{(j)}_a \partial_t  \psi^{(j)}_a - \frac{N}{2q} \iint_0^T \d t \d t' J^2(t,t') s_j s_{j'} \left( \frac{1}{N} \psi_a^{(j)}(t) \psi_a^{(j')}(t') \right)^q.
	\eeqn

At this stage, it is convenient to introduce the collective $(G,\Sigma)$ variables. Since we have two sets of fermions corresponding to $A_1$ and $A_2$, we introduce two collective fields 
	\begin{equation}
		G^{(1)}_{ij}(t,t') = \frac{1}{N_1}  \sum_{a = 1}^{N_1} \psi_a^{(i)}(t) \psi_a^{(j)}(t'),  \quad  	G^{(2)}_{ij}(t,t') = \frac{1}{N_2}  \sum_{a = N_1+1}^{N} \psi_a^{(i)}(t), \psi_a^{(j)}(t')  ,
	\end{equation}
and the corresponding Lagrange multipliers $\Sigma^{(1)}_{ij}(t,t')$ and $\Sigma^{(2)}_{ij}(t,t')$ to impose the constraints. We can now integrate out the fermions. The action in terms of the collective variables is 
 \begin{equation}
 \begin{split}
 	-\frac{I}{N} = \lambda &\log \Pf(\partial_t - \Sigma^{(1)})  + (1 - \lambda) \log \Pf( \partial_t - \Sigma^{(2)})  \\ 
 	& -\frac{1}{2}\iint_0^T \d t \d t' \left[ \lambda \Sigma_{ij}^{(1)}(t,t') G^{(1)}_{ij}(t,t') + 
 	(1 - \lambda) \Sigma_{ij}^{(2)}(t,t') G^{(2)}_{ij}(t,t') \right] \\ 
 	& + \frac{1}{2q} \iint_0^T
 \d t \d t' J^2(t,t') s_j s_{j'} G_{jj'}(t,t')^q,	\end{split}
 \label{eq:typaction}
 \end{equation}
 where $\Pf$ is the Pfaffian, and we have defined
 \begin{equation}
 	G_{ij}(t,t') = \lambda G^{(1)}_{ij} (t,t') + (1 - \lambda) G^{(2)}_{ij} (t,t').
 \end{equation}
 In the large $N$ limit, the path integral over the collective variables can be performed in the saddle point approximation. The equations of motion corresponding to the above action are:
 \beq
 \pa_{t}G^{(\alpha)}_{jj'}(t,t')-\Sigma_{jk}^{(\ka)}\star G^{(\ka)}_{kj'}(t,t')= \delta(t-t')\delta_{jj'},
 \eeq
 \beq
 \Sigma_{jj'}^{(\ka)}= s_js_{j'}J^2(t,t'){G_{jj'}^{(\ka)}(t,t')}^{q-1},
 \eeq
where $\alpha = 1$ corresponds to the fermions in $A_1$ while $\alpha=2$ corresponds to the fermions in $A_2$, the repeated $k$ index is summed, and the star product between two bi-local fields is defined as
\beq
(A\star B)(t,t') = \int \d t''\,A(t,t'')B(t'',t).
\eeq
Using the fact that $G^{(\ka)}$ and $\Sigma^{(\ka)}$ are both anti-symmetric, we can rewrite these equations in a more convenient form:
\beq
\left(\pa_{t}+\pa_{t'}\right)G^{(\alpha)}_{jj'}=\left(\Sigma_{jk}^{(\ka)}\star G^{(\ka)}_{kj'}-G_{jk}^{(\ka)}\star \Sigma^{(\ka)}_{kj'}\right),
\eeq
\beq
\Sigma_{jj'}^{(\ka)}= s_js_{j'}J^2(t,t'){G_{jj'}^{(\ka)}}^{q-1}.
 \eeq

Now, a  simplification happens in the Brownian SYK model --- recall that for Brownian SYK, $J^2(t,t') = J \delta(t-t')$. As a result, $\Sigma$ is ``diagonal'' (in time), and only the diagonal components of all the collective variables are relevant; the off-diagonal components drop out of the equations of motion. In fact, it is easy to see from the action that for Brownian SYK, the off-diagonal modes do not have any interesting dynamics and can be integrated out of the full path integral trivially \cite{Stanford:2021bhl}. 

Let us denote the diagonal components of the collective variables as 
\begin{equation}
	\begin{split}
		&G^{(\alpha)}_{ij}(t,t) =
		g_{ij}^{(\alpha)}(t) , \quad \Sigma^{(\alpha)}_{ij}(t,t') = \delta(t-t') \sigma^{(\alpha)}_{ij}(t) \,,
	\end{split}
\end{equation}	
The resulting equations of motion for the $(g,\sigma)$ variables are 
\begin{equation}
		\frac{\d g}{\d t}^{(\ka)} = \left[\sigma^{(\ka)}(t), g^{(\ka)}(t)\right], \quad \sigma^{(\ka)}_{ij}(t) = \sigma_{ij}(t) \equiv \begin{cases}
			J s_i s_j (g_{ij}(t))^{q-1}, & i \neq j \\
			0, & i = j
		\end{cases}
		\label{eq:typeom}
\end{equation}
where the equation on the left is written for the $g$ and $\sigma$ matrices and 
\begin{equation}
	g(t) = \lambda g^{(1)}(t) + (1 - \lambda) g^{(2)}(t) .
	\label{eq:typg}
\end{equation} 
In terms of the $(x_\ka,y_\ka,z_\ka)$ variables, we get the following equations of motion: 
\begin{equation}
	\begin{split}
		&\dot{x}_\ka = \frac{J}{2^{q-2}}( y^{q-1} z_\ka -  z^{q-1} y_\ka) \\
	&\dot{y}_\ka = \frac{J}{2^{q-2}} (x^{q-1}z_\ka -  z^{q-1}x_\ka) \\
	&\dot{z}_\ka = \frac{J}{2^{q-2}} (x^{q-1} y_\ka -  y^{q-1}x_\ka), \\ 
	\end{split}
	\label{eq:typeqx1}
\end{equation}
where recall that $x=\lambda x_1+ (1-\lambda)x_2$, and so on. These relations imply the following equations of motion for the total variables:
\begin{equation}
	\begin{split}
		&\dot{x} = \frac{J}{2^{q-2}}(y^{q-1} z -  z^{q-1} y) \\
		&\dot{y} = \frac{J}{2^{q-2}} ( x^{q-1}z -  z^{q-1} x) \\
		&\dot{z} = \frac{J}{2^{q-2}} (x^{q-1} y -  y^{q-1}x). \\
	\end{split}
\end{equation}

As an aside, the above equations of motion have a Hamiltonian structure. To see this, let us denote $\bx_{\ka} = (x_\ka,y_\ka,z_\ka)$ and $\bx = \lambda \bx_1 +(1-\lambda)\bx_2= (x,y,z)$. Further, we define $h_2(\bx) = \frac{1}{2}(x^2 - y^2 +z^2)$ and $h_q(\bx) = \frac{1}{q}(x^q-y^q+z^q)$. Then, equations \eqref{eq:typeqx1} take the succinct form
\beq
\dot{\bx}_{\ka}^I = \omega_{\ka}^{IJ}\frac{\pa h_q(\bx)}{\pa \bx^J_{\ka}},\;\;\;\omega^{IJ}_{\ka}=- p_{\ka} \epsilon^{IJK}\frac{\pa h_2(\bx_{\ka})}{\pa \bx_{\ka}^K},
\eeq
where $p_{\ka} = (\frac{1}{\lambda},\frac{1}{1-\lambda})$. Similarly, the equations for the total variables take the form
\beq
\dot{\bx}^I =\omega^{IJ}\frac{\pa h_q(\bx)}{\pa \bx^J},\;\;\; \omega^{IJ}= -\epsilon^{IJK}\frac{\pa h_2(\bx)}{\pa \bx^K}.
\eeq
Thus, these equations take the form of Hamilton's equations of motion -- the underlying phase space is that of two copies, labelled by $\ka$, of a co-adjoint orbit of $\mathfrak{sl}(2,\mathbb{R})$ specified by a constant value of the conserved quantity $h_2(\bx)$.\footnote{The solutions we will find turn out to have $2h_2(\bx) = 1$, so for the total variables $\bx$ the orbits in question are related to the ``continuous series'' of unitary $\mathfrak{sl}(2,\mathbb{R})$ representations by geometric quantization.  Understanding the significance of this structure is an interesting problem in its own right, but we will not address it further in this work.}
The Hamiltonian $h_q(\bx)$ couples the two copies, with the effective coupling constant being $\lambda$.
It may seem unusual that we have an odd number of variables (e.g. $(x,y,z)$) in Hamiltonian mechanics, but this is simply because we have parametrized the two-dimensional dynamics on the hypersurface $h_2(\bx) = $ const. in terms of coordinates in the ambient $\mathbb{R}^3$.

\subsubsection*{Normalization}
The fermionic path integral depicted in Figure~\ref{fig:contour2} is the result of writing $\Tr \sigma_L^2$ in a Hilbert space form and replacing the maximally entangled state projectors with connections in the contour.
However, translating a Hilbert space expression into a fermionic path integral comes with a standard normalization issue since Majorana fermions have a somewhat unusual Hilbert space interpretation (when they admit one at all).
So, we must relate the path integral $Z(T)$ to $\Tr \sigma_L^2(T)$ with an overall normalization that ensures $\Tr \sigma_L^2(T=0) = 1$.

When $T=0$, the contour in Figure~\ref{fig:contour2} consists of four disconnected circles which are not coupled by any time evolution.
The result of the path integral for a single free Majorana fermion on a circular contour of length $T$ with antiperiodic boundary conditions is actually equal to $\sqrt{2}$, independent of $T$, so in the limit $T \to 0$ we still have $\sqrt{2}$.
As such, when $T=0$, the contour in Figure~\ref{fig:contour2} yields $Z(T=0) = 2^N$ since there are $N_1$ Majorana fermions on the two $A_1$ circles and $N_2$ Majorana fermions on the two $A_2$ circles.
So we must relate the path integral $Z(T)$ to $\Tr \sigma_L^2(T)$ by the formula
\begin{equation}
    \Tr \sigma_L^2(T) = 2^{-N} Z(T) .
\label{eq:normalization}
\end{equation}

\subsection{Solutions: qualitative discussion}

We will first qualitatively discuss what the solutions to the equations of motion \eqref{eq:typeqx1} should look like, leaving a quantitative treatment for Sections~\ref{sec:perturbative-disk-soln},~\ref{sec:wormhole-soln}, and~\ref{sec:lattice-solns}.
When $N_1=0$ (i.e., $\lambda =0$), then the $\bx_2$ equations are easy to solve. In this case, the boundary conditions imply that the solution stays at the fixed point $\bx^*_2=(0,0,1)$.
When $\lambda$ is small but non-zero, we expect that this saddle point remains, but with small corrections. In particular, the $\bx_2$ variables will stay close to their original fixed point values. The corrections to the $\bx_2$ solutions can be obtained in perturbation theory in $\lambda$, and we describe them in detail in Section~\ref{sec:perturbative-disk-soln}. (Recall from the discussion under \eqref{eq:classical-limit} that $\lambda \ll 1$.)
This resulting solution is the dominant saddle point at small times, and is the analogue of the ``disconnected'' contribution in equation \eqref{eq:Haaraverage}, or the disconnected geometry in JT gravity \cite{Balasubramanian:2022fiy}.

When $\lambda$ is small, to zeroth order, the solution for $\bx_2$ variables will be unaffected by the $\bx_1$ variables, and in particular will correspond to a fixed point of the Hamiltonian picked out by the boundary conditions as we have just described. 
However, the initial backreaction on the $\bx_1$ variables will be large.
The source of this strong backreaction is the mismatch between the boundary conditions of the $\bx_1$ and $\bx_2$ variables.
As the $A_1$ system is small compared to $A_2$, the Brownian dynamics quickly thermalizes the $A_1$ system so that the correlation between contours (in Figure~\ref{fig:contour2})  can be measured with any subset of all $N$ fermions; at the level of the solutions, this means that we will have $\bx_1 \approx \bx_2$ at all times except in small neighborhoods around $t=0$ and $t=T$ where we expect large transient behaviors for $\bx_1$ to arrive at their ``thermalized'' values.
These transient behaviors can be computed analytically in perturbation theory for the disconnected solution and must be treated numerically otherwise.
These qualitative properties of the $\bx_1$ solutions hold both for the disconnected solution in Section~\ref{sec:perturbative-disk-soln} as well as the other solutions we now describe.

In addition to generating nontrivial time-dependence for the disconnected solution, turning on a small $\lambda$ has another important effect -- it gives rise to new ``tunneling'' solutions (i.e., instantons) which are absent at $\lambda =0$. 
The tunneling allows the $\bx_2$ variables to jump between different fixed points; the leading tunneling solution jumps from $\bx_2(0) \approx (1,-1,1)$ to $\bx_2(T) \approx (1,1,1)$.
The $\bx_1$ variables again have large transient signals near $t=0$ and $t=T$, but this time both transients are different than the ones which occur for the disconnected solution due to the different fixed points approached by the $\bx_2$ variables, and in fact these are the only other two types of transient behavior which can occur.
This saddle point, which we describe in Section~\ref{sec:wormhole-soln}, is the analogue of the ``connected'' saddle point in equation \eqref{eq:Haaraverage}, or the ``wormhole'' in JT gravity \cite{Balasubramanian:2022fiy}. While it is suppressed by a factor of $e^{-(1-2\lambda)N}$ relative to the leading, disconnected solution, the contribution of the disconnected saddle point decays exponentially in time. 
So, at a time $t_* \sim O(N)$, there is an exchange of dominance between these two saddle points.

There are also other tunneling saddle points, described in Section~\ref{sec:lattice-solns}, where the $\bx_2$ variables tunnel back and forth multiple times between the two possible initial fixed points $\bx_2 \approx (0,0,1)$ and $\bx_2 \approx (1,-1,1)$ and the two possible final fixed points $\bx_2 \approx (0,0,1)$ and $\bx_2 \approx (1,1,1)$; these are even more subleading in powers of $e^{-N}$, and occur with all possible combinations of the previously described types of transient behaviors for the $\bx_1$ variables.
Explicitly, there are two possible behaviors at $t=0$ and two at $t=T$ corresponding to the possible initial and final fixed points for $\bx_2$, and all four combinations of initial and final transient behaviors occur in the multiply tunneling solutions.
These multiply tunneling solutions show interesting behavior as a function of $T$. We will see that they become genuine solutions of the equations \eqref{eq:typeqx1} only after certain critical values of $T$, related to the scrambling time.
Before these critical times, these configurations are actually off-shell.
Configurations which tunnel more times take longer to become solutions.
As the presence of these contributions is important for unitarity of the overall evolution \cite{Stanford:2021bhl}, it is intriguing that they can be invisible on-shell for a parametrically (though not polynomially) long time in $N$.

In summary, we began with the goal of studying the error correction dynamics of a family of unitary operators with increasing average circuit complexity.
The specific family which we chose for convenience was the set of time evolution operators in the Brownian SYK model, a family of time-dependent Hamiltonians which are essentially a continuous random circuit.
We found that the error correction dynamics are governed by the quantity $\cF_{\Psi'}(\re:\en)$, and this quantity in turn depends on a Brownian SYK path integral (Figure~\ref{fig:contour2}).
What we have just discussed are the saddle point solutions to that path integral. Evaluating the effective action of these solutions will allow us to draw conclusions about the error correction behavior of the family of unitary operators with increasing complexity.

\subsection{Disconnected solution}\label{sec:perturbative-disk-soln}
We will first solve for the disconnected solution at small, non-zero $\lambda$, and evaluate its on-shell action together with the one-loop determinant.
We begin by expanding our variables in a power series expansion in $\lambda$: 
\beq
\bx_{\ka}  = \sum_{n=0}^{\infty} \lambda^n \bx^{(n)}_{\ka},\;\;\;\bx  = \sum_{n=0}^{\infty} \lambda^n \bx^{(n)}.
\eeq
At $O(\lambda^0)$, we must have $\bx^{(0)}=\bx_2^{(0)}$. Therefore, the boundary conditions, equations (\ref{eq:typbc3}) and (\ref{eq:typbc4}), imply that at leading order these variables sit at a fixed point of the Hamiltonian: 
\beq
\bx^{(0)}(t) = \bx_2^{(0)}(t) = (0,0,1).
\eeq
After substituting these solutions in  (\ref{eq:typeqx1}), we get the following equations for $\bx_1^{(0)}$:
\begin{equation}
\begin{split}
	&\dot{x_1}^{(0)} = - \frac{J}{2^{q-2}} \,y_1^{(0)}, \\
	&\dot{y_1}^{(0)} = - \frac{J}{2^{q-2}} \,x_1^{(0)}, \\
	&\dot{z_1}^{(0)} = 0.
	\end{split}
\end{equation}
We need to solve these equations with the boundary conditions (\ref{eq:typbc1}) and (\ref{eq:typbc3}). The solution is
\begin{equation}
\begin{split}
	&x_1^{(0)}(t) = \frac{\cosh\left(\frac{J}{2^{q-2}} (t - \frac{T}{2})\right)}{\cosh\left({\frac{JT}{2^{q-1}}}\right)} , \\ &y_1^{(0)}(t) = -\frac{\sinh\left(\frac{J}{2^{q-2}} (t - \frac{T}{2})\right)}{\cosh\left({\frac{JT}{2^{q-1}}}\right)} ,
	\\ &z_1^{(0)}(t) = \tanh\left(\frac{JT}{2^{q-1}}\right).
\end{split}
\label{eq:x1-backreaction}
\end{equation}
We can think of this solution as the backreaction of the $\bx_2^{(0)}$ variables on the $\bx_1^{(0)}$ variables. For instance, $x_1$ starts off at one at $t=0$, but after a brief transient behavior, it thermalizes to a small value of about $e^{-\frac{JT}{2^{q-1}}}$ owing to its coupling to the $\bx_2$ variables, which act like a bath and dynamically force $\bx_1 \approx \bx_2$.
There is another transient near $t=T$, where $\bx_1$ again deviates from $\bx_2$ significantly to reach the final boundary condition.

At order $\lambda$, we note from $\bx = \lambda \bx_1 + (1-\lambda)\bx_2$, that 
\beq
\bx^{(1)} = \bx^{(1)}_2 +\bx^{(0)}_1 - \bx^{(0)}_2.
\eeq
Now we use the $O(\lambda^0)$ solutions to find the following boundary conditions up to $O(\lambda)$:
\begin{equation}
\begin{split}
&x_1^{(1)}(0) = 0, \quad y_1^{(1)}(0) = z_1^{(1)}(0) ,\\
&x_1^{(1)}(T) = 0, \quad y_1^{(1)}(T) = -z_1^{(1)}(T)  ,
\end{split}
\end{equation}
\begin{equation}
\begin{split}
&x_2^{(1)}(0) = - y_2^{(1)}(0), \quad z_2^{(1)}(0) = 0  ,\\
&x_2^{(1)}(T) = y_2^{(1)}(T), \quad z_2^{(1)}(T) = 0 ,
\end{split}
\end{equation}
\begin{equation}
	\begin{split}
		&\bx^{(1)}(0) = \left(1 + x^{(1)}_2(0),\tanh\left(\frac{JT}{2^{q-1}}\right) + y^{(1)}_2(0), \tanh\left(\frac{JT}{2^{q-1}}\right) - 1\right) ,\\
		&\bx^{(1)}(T) = \left(1 + x^{(1)}_2(T),-\tanh\left(\frac{JT}{2^{q-1}}\right) + y^{(1)}_2(T), \tanh\left(\frac{JT}{2^{q-1}}\right) - 1\right) .
	\end{split}
\end{equation}
With these boundary conditions in hand, we can in principle solve for all the variables at $O(\lambda)$. However, in what follows, we will only need $\bx^{(1)}$ in order to evaluate the on-shell action up to $O(\lambda)$. The corresponding differential equations are given by
\begin{equation}
\begin{split}
	&\dot{x}^{(1)} = - \frac{J}{2^{q-1}} \,y^{(1)}, \\
	&\dot{y}^{(1)} = - \frac{J}{2^{q-1}} \,x^{(1)}, \\
	&\dot{z}^{(1)} = 0.
	\end{split}
\end{equation}
The solution is:
\begin{equation}
	\begin{split}
		&x^{(1)}(t) = \left(1 + \tanh\left(\frac{JT}{2^{q-1}}\right) \right) \cosh \left(\frac{J}{2^{q-2}} \left(t - \frac{T}{2}\right)\right) e^{-\frac{JT}{2^{q-1}}} , \\
		&y^{(1)}(t) = -\left(1 + \tanh\left(\frac{JT}{2^{q-1}}\right) \right) \sinh \left(\frac{J}{2^{q-2}} \left(t - \frac{T}{2}\right)\right) e^{-\frac{JT}{2^{q-1}}} , \\
		&z^{(1)}(t) = \tanh\left(\frac{JT}{2^{q-1}}\right) -1 .
	\end{split}
\end{equation}
\subsubsection*{Classical on-shell action}
Having obtained the classical solutions, we can evaluate the action in (\ref{eq:typaction}) at leading order in $\lambda$.
We first compute the Pfaffian for the $A_1$ fermions:
\begin{equation}
\Pf(\partial_t - \sigma^{(1)}) = 
\stretchint{7ex}_{\mkern-12mu
	\psi^{(1)} = i \psi^{(2)},\ 
	\psi^{(3)} = i \psi^{(4)}
}^{
	\psi^{(1)} = -i\psi^{(2)},\ 
	\psi^{(3)}=-i\psi^{(4)}
} \hspace{-3cm} \mathcal{D}\psi^{(1)} \dots \mathcal{D}\psi^{(4)} \exp \left( -\frac{1}{2} \int_0^T \d t \left[ \psi^{(j)} \partial_t \psi^{(j)} - \sigma_{jj'}(t) \psi^{(j)} \psi^{(j')} \right] \right) .
\end{equation}
On the right hand side of the above expression, we have used the fact that on-shell, $\sigma^{(1)}_{ij}(t) = \sigma_{ij}(t)$; see equation \eqref{eq:typeom}. Following \cite{Stanford:2021bhl}, we can write the Pfaffian in a Hilbert space representation in terms of a single qubit:\footnote{Naively, we would need two qubits given that there are four Majorana fermions. However, since time evolution preserves fermion number, we need only use one qubit.}
\begin{equation}
	\Pf \left(\partial_t - \sigma^{(1)}\right) = 2\times \bra{+} \mathcal{T}\exp\left(-\int_0^T \d t\, h(t)\right) \ket{+} ,
\label{eq:pfaffian-A1}
\end{equation}
where $h(t)$ is the qubit Hamiltonian:\footnote{The factor of 2 appearing in \eqref{eq:pfaffian-A1} ensures that, when $T=0$, the Pfaffian gives 2, as this is the result for a single Majorana fermion path integral on two disjoint circles of any length with antiperiodic boundary conditions. }
	\begin{equation}
	h(t) = \frac{J}{2^{q-1}} \left( -x^{q-1}(t) X + i y^{q-1}(t) Y - z^{q-1}(t) Z \right) ,
	\end{equation}
and $X,Y,Z$ are the Pauli matrices. Further, $|+\rangle$ is an eigenstate of the Pauli $X$ operator with the eigenvalue +1: $X|+\rangle = |+\rangle$. The initial and final states are fixed by the boundary conditions of the path integral. Since the leading contribution to $x^{q-1}(t),y^{q-1}(t)$ is at $O(\lambda^{q-1})$ we can ignore them in the evaluation of the Pfaffian (assuming $q\geq 4$). The Pfaffian is therefore given by 
\begin{equation}
\begin{split}
\Pf \left(\partial_t - \sigma^{(1)}\right) &= 2\times\bra{+} \exp\left(\int_0^T \d t\, \frac{J}{2^{q-1}} z^{q-1}(t) Z\right) \ket{+} + O(\lambda^{q-1})\\
&= 2\cosh \left(\frac{JT}{2^{q-1}} z^{q-1}(0) \right) + O(\lambda^{2}) .
\end{split}
	\label{eq:typpf1}
\end{equation}
In the second step, we used the fact that $z(t)$ is a constant at $O(\lambda)$. Similarly, we can evaluate the Pfaffian for the $A_2$ fermions:
\begin{equation}
\begin{split}
\Pf(\partial_t - \sigma^{(2)}) &= 
\stretchint{7ex}_{\mkern-12mu
	\psi^{(1)} = i \psi^{(4)},\ 
	\psi^{(2)} = i \psi^{(3)}
}^{
	\psi^{(1)} = -i\psi^{(4)},\ 
	\psi^{(2)}=-i\psi^{(3)}
} \hspace{-3cm} \mathcal{D}\psi^{(1)} \dots \mathcal{D}\psi^{(4)} \exp \left( -\frac{1}{2} \int_0^T \d t \left[ \psi^{(j)} \partial_t \psi^{(j)} - \sigma_{jj'}(t) \psi^{(j)} \psi^{(j')} \right] \right) \\& = 2\times \bra{0} \exp{\left(\frac{JT}{2^{q-1}}z^{q-1}(0)Z\right)}\ket{0} + O(\lambda^{q-1}) \\ &=  2\exp{\left(\frac{JT}{2^{q-1}}z^{q-1}(0)\right)} + O(\lambda^{q-1}) . \end{split} \label{eq:typpf2}\end{equation}
Here, $\ket{0}$ is an eigenstate of the Pauli $Z$ operator with eigenvalue +1: $Z \ket{0} = \ket{0}$. 

\begin{figure}
    \centering
    \includegraphics[width=.49\textwidth]{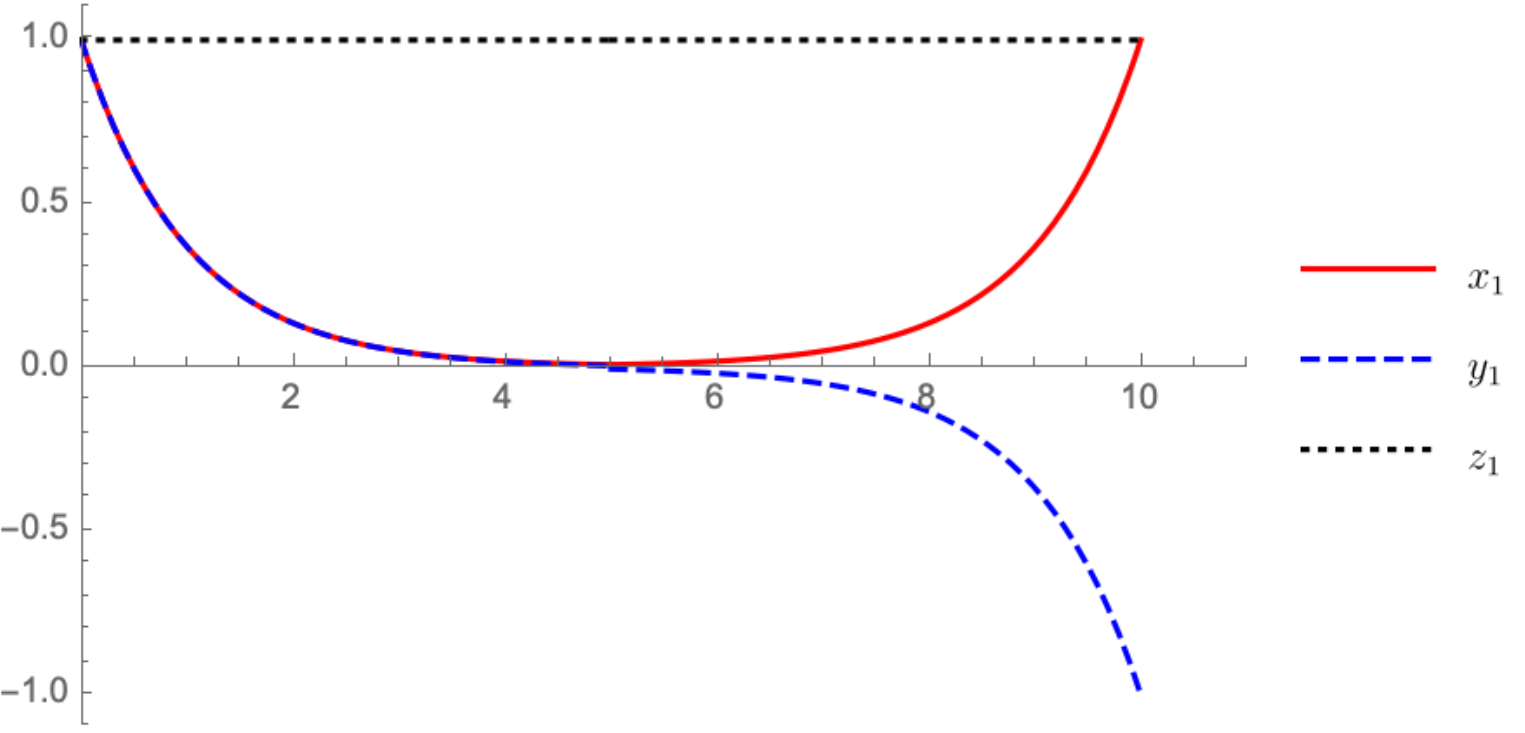}
    \includegraphics[width=.49\textwidth]{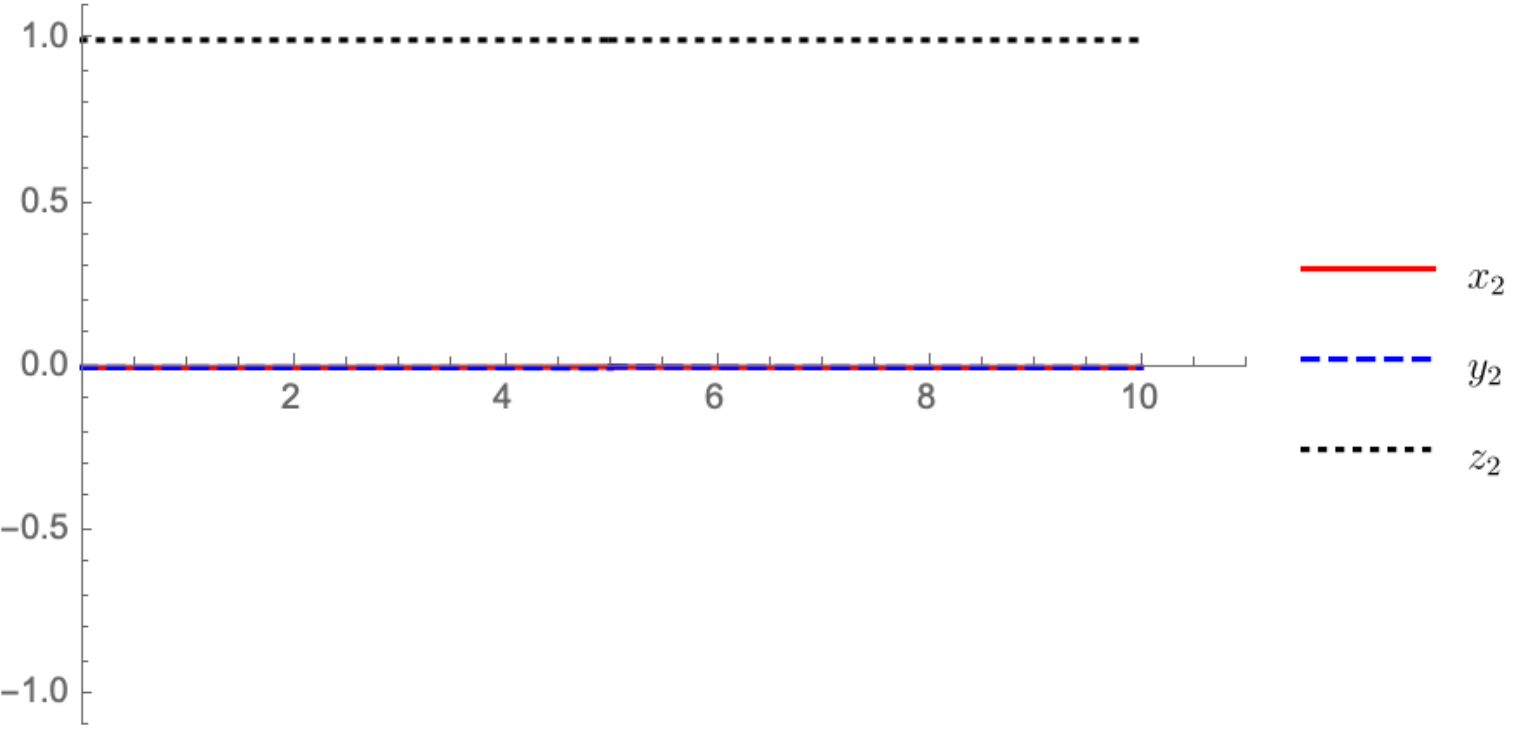}
    \caption{\small{The disconnected solution for the $\bx_1$ (left) and $\bx_2$ (right) variables with $J=q=4$, $r = 1-10^{-8}$, and $\lambda = 10^{-4}$.  There is a long region where $\bx_1 \approx \bx_2$ around the fixed point $(0,0,1)$, while there are large transient behaviors where $\bx_1 \neq \bx_2$ near $t=0$ and $t=T\approx 10$ due to the difference between the $\bx_1$ and $\bx_2$ boundary conditions.  The difference between the exact numerical $\bx_1$ solution shown here and the perturbative solution \eqref{eq:x1-backreaction} is essentially invisible. }}
    \label{fig:disk-soln}
\end{figure}

Having  evaluated the Pfaffians to $O(\lambda)$, we now evaluate the rest of the terms in the action  following \cite{Stanford:2021bhl}:
\begin{equation}
	\begin{split}
		-\frac{1}{2}\iint_0^T \d t \d t' &\left[ \lambda \Sigma_{ij}^{(1)}(t,t') G^{(1)}_{ij}(t,t') + 
		(1 - \lambda) \Sigma_{ij}^{(2)}(t,t') G^{(2)}_{ij}(t,t') \right] \\ 
		& \quad + \frac{1}{2q} \iint_0^T
		\d t \d t' J^2(t,t') s_j s_{j'} G_{jj'}(t,t')^q  \\ 
		&\quad = -\frac{JT}{2^{q-1}q} -\frac{1}{2}\int_0^T \d t \, \left[ \sigma_{ij}(t) \, g_{ij}(t)  - \frac{1}{q}
		 \, s_j s_{j'} \, g_{jj'}(t,t')^q \right] \\ 
		&\quad = -\frac{JT}{2^{q-1}q} - \frac{JT}{2^{q-1}}\frac{q-1}{q}r \\
		&\quad = \frac{JT}{2^{q-1}q}(r-1) - \frac{JT}{2^{q-1}}r ,
	\end{split}
\end{equation}
where $r = qh_q(\bx) = x^q(t) - y^q(t) + z^q(t)$ is a constant of motion, namely the total energy. 
Note that this form of the bulk contribution to the action is independent of the particular solution we are considering.
Now, we can combine the above terms with equations (\ref{eq:typpf1}) and (\ref{eq:typpf2}) to obtain the full on-shell action for the disconnected saddle point:
\begin{equation}
	\begin{split}
		\frac{-I}{N} &= \log 2 + \lambda \log \cosh \left(\frac{JT}{2^{q-1}} \, z^{q-1}(0) \right) + (1 - \lambda) \frac{JT}{2^{q-1}} \, z^{q-1}(0) +  \frac{JT}{2^{q-1}q}(r-1) - \frac{JT}{2^{q-1}}r \\
	&=  \log 2 + \lambda \log \cosh \left(\frac{JT}{2^{q-1}} \right) + \frac{JT}{2^{q-1}}\left( (1 - \lambda)z^{q-1}(0) + \frac{r-1}{q} -r \right)  \\
	&=  \log 2 + \lambda \left( \log \cosh \left(\frac{JT}{2^{q-1}} \right) - \frac{JT}{2^{q-1}} \right) + O(\lambda^2) .
	\end{split} 
\end{equation}
As the normalization relation \eqref{eq:normalization} cancels the leading $\log 2$ in the effective action, the contribution of the disconnected saddle point in the large $N$ limit is given by
\begin{equation}
	 \Tr{\sigma_L^2}\Big|_{\text{disc}} \approx \left( \frac{1 + \exp \left( -\frac{JT}{2^{q-2}}\right)}{2} \right)^{N\lambda} .
\end{equation}
This formula is consistent with physical expectations; see the discussion around equation \eqref{eq:expec}. At small times $\frac{JT}{2^{q-1}} \ll 1$, we find that $\Tr \sigma_L^2 \rightarrow 1$. This is expected, since in this limit, $U_A(T)$ does not introduce much entanglement between $L$ and $A_1$. On the other hand, at late times $\frac{JT}{2^{q-1}} \gg 1$,  $\Tr \sigma_L^2 \rightarrow 2^{-N \lambda} = \frac{1}{d_\co^2}$, as we anticipated based on monogamy of entanglement. As a further check, we also reproduce the above formula from a ``Hamiltonian'' point of view in Appendix \ref{sec:Hamiltonian}. For now, we proceed to evaluate the one-loop determinant around the disconnected solution. 
But, before doing so, we display the numerical solutions for $\bx_1$ and $\bx_2$ in Figure~\ref{fig:disk-soln}, where we can clearly see the leading order nontrivial time-dependence of $\bx_1$ takes the rough hyperbolic forms we found for $\bx_1^{(0)}(t)$ in \eqref{eq:x1-backreaction}.

\subsubsection*{One-loop determinant}

To compute the one-loop determinant in the path integral formalism, we need to expand the action around the saddle point and integrate over small fluctuations. We will follow the notations and conventions of Appendix B in \cite{Stanford:2021bhl}. Recall that the action is given by
	\begin{equation}
\begin{split}
\frac{-I}{N} = \lambda &\log \Pf(\partial_t - \sigma^{(1)})  + (1 - \lambda) \log \Pf( \partial_t - \sigma^{(2)})  \\ 
& -\frac{1}{2}\int_0^T \d t \left[ \lambda \sigma_{ij}^{(1)}(t) g^{(1)}_{ij}(t) + 
(1 - \lambda) \sigma_{ij}^{(2)}(t) g^{(2)}_{ij}(t) \right] \\ 
& + \frac{J}{2q} \int_0^T
\d t \, s_j s_{j'} \, g_{jj'}(t)^q .	\end{split}
\label{eq:pth-det-act}
\end{equation}
We can write the Pfaffian in the Hilbert space representation, as in equations (\ref{eq:typpf1}) and (\ref{eq:typpf2}):
\begin{equation}
\begin{split}	
	-\frac{I}{N} = \log 2 + \lambda  &\log \left[ \bra{+} \exp \left( \int_0^T \d t \left( -\sigma_x^{(1)}(t) X + i\sigma_y^{(1)}(t) Y - \sigma_z^{(1)}(t)Z\right)\right) \ket{+}\right]\\  &+ (1- \lambda) \log \left[ \bra{0} \exp \left( \int_0^T \d t \left( -\sigma_x^{(2)}(t) X + i\sigma_y^{(2)}(t) Y - \sigma_z^{(2)}(t)Z\right)\right) \ket{0}\right] \\&+ \lambda \int_0^T \d t \left[\sigma_x^{(1)}(t)x^{(1)}(t)-\sigma_y^{(1)}(t)y^{(1)}(t)+\sigma_z^{(1)}(t)z^{(1)}(t)\right] \\& + (1- \lambda)\int_0^T \d t \left[\sigma_x^{(2)}(t)x^{(2)}(t)-\sigma_y^{(2)}(t)y^{(2)}(t)+\sigma_z^{(2)}(t)z^{(2)}(t)\right] \\ &  -\frac{J}{2^{q-1}q} \int_0^T \d t  \left( 1 - x^q(t)+ y^{q} (t) - z^q(t) \right) .
\end{split}
	\end{equation}
We now expand around the saddle point solution $\bx^{(\ka)}_*$ found in Section~\ref{sec:perturbative-disk-soln}. We will use hatted variables to denote fluctuations:
\beq
\bx^{(\ka)}= \bx^{(\ka)}_* +\sqrt{\hbar^{(\ka)}} \,\hat{g}^{(\ka)},\;\;\boldsymbol{\sigma}^{(\ka)} = \boldsymbol{\sigma}^{(\ka)}_* + \sqrt{\hbar^{(\ka)}}\,\hat{\sigma}^{(\ka)},
\eeq
\begin{equation}
	\begin{split}
		\hat{\sigma}^{(\ka)} = \left(\hat{\sigma}^{(\ka)}_x,  \hat{\sigma}^{(\ka)}_y,\hat{\sigma}^{(\ka)}_z\right) , \quad  \hat{g}^{(\ka)} = \left(\hat{x}^{(\ka)},  \hat{y}^{(k)},\hat{z}^{(\ka)}\right),
	\end{split}
\end{equation}
where $\hbar^{(\ka)} = \left(\frac{1}{N\lambda}, \frac{1}{N(1-\lambda)}\right)$. The quadratic action for the fluctuations has the following form: 
\begin{equation}
	-\hat{I} = \begin{pmatrix} \hat{\sigma}^{(1)} \\ \hat{g}^{(1)} \\\hat{\sigma}^{(2)}\\\hat{g}^{(2)} \end{pmatrix}. M.
	 \begin{pmatrix} \hat{\sigma}^{(1)} \\ \hat{g}^{(1)} \\\hat{\sigma}^{(2)}\\\hat{g}^{(2)} \end{pmatrix}.
\end{equation}
where 
\begin{equation}
M = 	\begin{pmatrix}
 K_1 &  S & 0 & 0 \\
 S & \lambda \tilde{S} & 0 & \sqrt{\lambda(1 - \lambda)} \tilde{S} \\
0 & 0&  K_2 & S  \\
0 & \sqrt{\lambda(1 - \lambda)} \tilde{S}&   S & (1-\lambda) \tilde{S} 
\end{pmatrix} .
\end{equation}
 The matrices $K_1$ and $K_2$ can be derived by variation of the Pfaffian terms at quadratic order. $S$ and $\tilde{S}$ are defined as
\begin{equation}
	S = \delta(t_{12}) \begin{pmatrix}
		1 & 0 & 0 \\ 
		0 & -1 & 0 \\ 
		0 & 0 & 1 
	\end{pmatrix}, \quad \tilde{S} =  \frac{J(q-1)}{2^{q-2}}\delta(t_{12}) \begin{pmatrix}
	0 & 0 & 0 \\ 
	0 & 0 & 0 \\ 
	0 & 0 & 1+ O(\lambda) 
	\end{pmatrix} .
\end{equation}
To compute the determinant at leading order in $\lambda$, it turns out to be sufficient to note that $K_2$ has non-zero matrix elements only in $\hat{\sigma}^{(2)}_x$ and $\hat{\sigma}^{(2)}_y$. Therefore, $K_2$ satisfies the relation 
 \begin{equation}
 \tilde{S} K_2 = K_2 \tilde{S} = 0 . \label{eq:k2s}
 \end{equation}  Moreover, we will only need the $(\sigma^{(1)}_{z},\sigma^{(1)}_{z})$ component of $K_1$ which is
 \begin{equation}
 	\begin{split}
 		\iint_0^T \d t_1 \d t_2 \, \sigma^{(1)}_{z}(t_1) \, K_{1}^{zz}(t_1,t_2)\, \sigma^{(1)}_{z}(t_2)  &= \frac{1}{2} \sech^{2} \left(\frac{JT}{2^{q-2}}\right)  \left( \int_0^T  \d t \, \sigma_z^{(1)}(t) \right)^2 .
 	\end{split}
 \end{equation}
We can now compute the determinant of $M$ to leading order in $\lambda$:
\begin{equation}
	\det{M} = \det{A} \det{C} \left[1 - \lambda \Tr \left(A^{-1} B C^{-1} B \right)\right] + O(\lambda^2) ,
\end{equation}
where
\begin{equation}
	A = \begin{pmatrix}
		K_1 & S \\ S & \lambda \tilde{S}
	\end{pmatrix}, \quad B = \begin{pmatrix}
		0 & 0 \\ 0 & \tilde{S} \;
	\end{pmatrix}, \quad  C = \begin{pmatrix}
	K_2 & S \\ S & (1 - \lambda)\tilde{S}
	\end{pmatrix}.
\end{equation}
We first compute the trace term in the determinant
\begin{equation}
\begin{split}
		 \Tr \left(A^{-1} B C^{-1} B \right) &= \Tr \left[ A^{-1}_{22} \tilde{S} C^{-1}_{22} \tilde{S} \right]  \\&= \Tr \left[S K_1 S \tilde{S} S K_2 (S - \tilde{S} K_2)^{-1} \tilde{S} \right] + O(\lambda) \\ &\approx \Tr \left[ SK_1\tilde{S} K_2 \tilde{S} \right] \\ &= 0 .
\end{split}
\end{equation}
In the first step, we used the fact that $B_{22}$ is the only non-zero entry in $B$. In the second step, we inserted the $A_{22}^{-1}$ and $C_{22}^{-1}$ components upto $O(\lambda)$ corrections. The third and fourth step follow from the relation (\ref{eq:k2s}). Using relation (\ref{eq:k2s}) once again, we conclude that $\det C = 1$. We are left with the evaluation of $\det A$. 
\begin{equation}
	\begin{split}
		\det A &= \det A |_{\lambda = 0} \left[1 - \lambda \Tr \left(K_1 \tilde{S} \right)\right] \\ 
		&= 1 - \lambda \frac{J(q-1)}{2^{q-2}}\iint_0^T \d t_1 \d t_2  \delta(t_{12}) K^{zz}(t_1,t_2) \\
		&= 1 - \lambda \frac{JT(q-1)}{2^{q-1}}\sech^2\left(\frac{JT}{2^{q-1}}\right) .
	\end{split}
\end{equation}
Thus,
\begin{equation}
    \det M = 1 - \lambda \frac{JT(q-1)}{2^{q-1}}\sech^2 \left( \frac{JT}{2^{q-1}} \right) + O(\lambda^2).
\label{eq:one-loop-calc}\end{equation}
So, the one-loop determinant does not significantly modify the $T$ dependence of $\Tr \sigma_L^2$ at leading order.
The coefficient of the $O(\lambda)$ term above is bounded by an $O(q)$ number, and $q\lambda$ is always small in our regime of interest.

\subsection{Connected solution}\label{sec:wormhole-soln}

When $\lambda$ is slightly non-zero, $\bx_2 \approx (0,0,1)$ is not the only fixed point for the $\bx_2$ variables which enters the analysis.
The leading solution involving more than one fixed point is the tunneling solution between the $\bx_2 \approx (1,-1,1)$ and $\bx_2 \approx (1,1,1)$ fixed points.
This solution has nontrivial time-dependence for the $\bx_2$ variables which involves an initial region where $\bx_2(t) \approx (1,-1,1)$, then a transition to the $\bx_2(t) \approx (1,0,0)$ fixed point where the solution remains for a long period, and then a final transition to the $\bx_2(t) \approx (1,1,1)$ fixed point. 
The $\bx_1$ solution has large transient behaviors in the initial and final fixed point regions, but matches very closely with $\bx_2$ in the long region where $\bx_2 \approx \bx_1 \approx (1,0,0)$.

\begin{figure}
    \centering
    \includegraphics[width=0.49\textwidth]{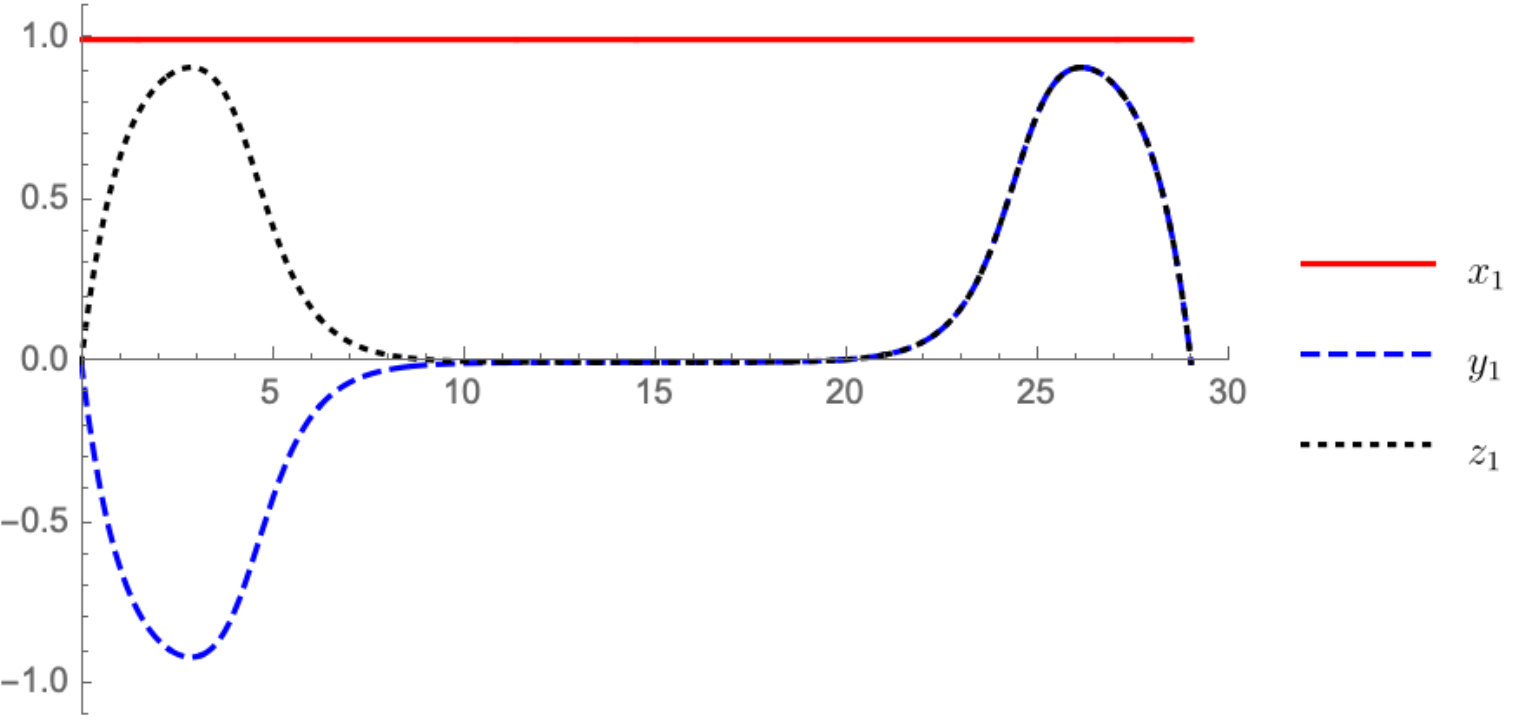}
    \includegraphics[width=0.49\textwidth]{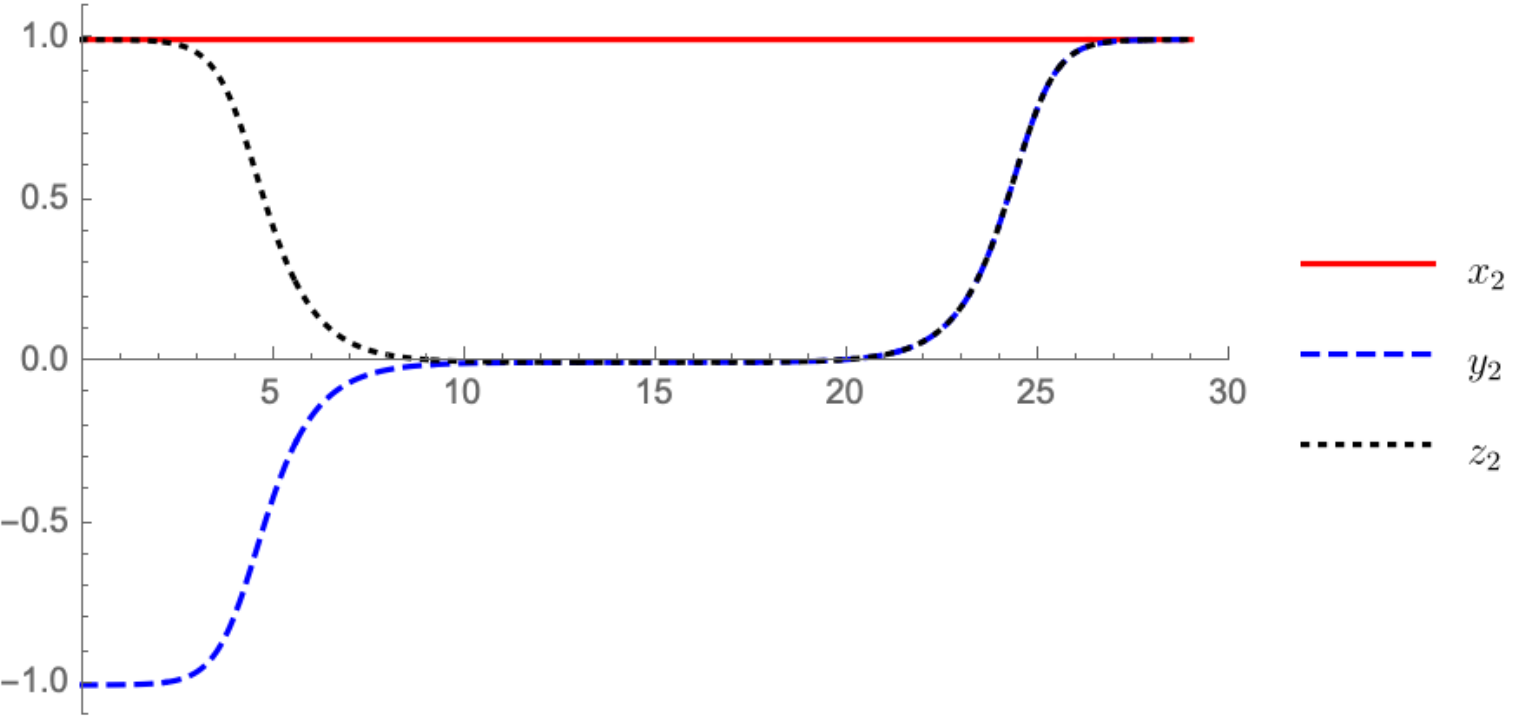}
    \caption{\small{The connected solution for the $\bx_1$ (left) and $\bx_2$ (right) variables with $J=q=4$, $r = 1-10^{-8}$, and $\lambda = 10^{-4}$.  There is a long region where $\bx_1 \approx \bx_2$ around the fixed point $(1,0,0)$, while there are large transient behaviors where $\bx_1 \neq \bx_2$ near $t=0$ and $t=T\approx 29$ due to the difference between the $\bx_1$ and $\bx_2$ boundary conditions.  The regions where $\bx_1$ displays transient behavior are comparable in size to a transition region where the $\bx_2$ variables tunnel between fixed points.}}
    \label{fig:connected-soln}
\end{figure}

Because this solution is non-perturbative in $\lambda$, we cannot hope to use perturbation theory to evaluate the effective action.
We will instead follow the approximate analysis of \cite{Stanford:2021bhl}.
Unlike the disconnected solution we described in Section~\ref{sec:perturbative-disk-soln}, the connected solution is suppressed exponentially in $N$, and the main aim of our approximate analysis will be to demonstrate this suppression quantitatively.
The numerical connected solution is shown in Figure~\ref{fig:connected-soln}. We will give an approximate analytical computation of the action for this solution.
As argued in Section~\ref{sec:perturbative-disk-soln}, for any solution of the equations of motion the bulk terms in the effective action contribute to the total path integral $\exp(-JT/2^{q-1})$ for $r \approx 1$. 
So what remains is to evaluate the two Pfaffian contributions, again using the qubit Hamiltonian approach.

We see in Figure~\ref{fig:connected-soln} that there is a long region with $\bx_1 \approx \bx_2 \approx (1,0,0)$. In this region, we may approximate the time-ordered exponential expressions as projectors $\ket{+}\bra{+}$, the lowest energy state of the Hamiltonian $-JTX/2^{q-1}$ generating the time evolution in that region.\footnote{When $T$ is on the order of $(1/J)\log N$ and not much larger, there are exponentially suppressed $T$ dependent corrections to this projector which lead to $O(1)$ factors in the wormhole contribution to the R\'enyi mutual information.  While these corrections could be addressed in the path integral formalism we employ here, it is easier to study them in the Hamiltonian picture (Appendix~\ref{sec:Hamiltonian}).  We will continue to approximate the long region as a projector because these corrections are highly subleading by the time the wormhole dominates at $T \sim N/J$ and are therefore unimportant for the qualitative error correction properties of the Brownian circuit.}
The energy contribution from this ground state exactly cancels the bulk term, so the result of the long region for both $\bx_1$ and $\bx_2$ is a projector $\ket{+}\bra{+}$.
At this point, the analysis splits between $\bx_1$ and $\bx_2$.

The $\bx_2$ variables include an initial region around the $(1,-1,1)$ fixed point and a final region around $(1,1,1)$.
Both of these regions share an important property with their adjacent transition regions, namely that the first has $y_2 \approx -z_2$ and the second has $y_2 \approx z_2$.
Because these regions are adjacent to the long middle region that yields a projector $\ket{+}\bra{+}$, we may use the null state relations $\bra{+}(iY+Z)= 0$ and $(iY-Z)\ket{+} = 0$ to conclude that the Pfaffian does not depend on the precise details of these transition regions nor on the initial and final fixed point regions, and the remaining $X$ term in the Hamiltonian simply cancels against the bulk contribution as was the case in the long middle fixed point region.
So, the overall Pfaffian for the $\bx_2$ variables is determined by the overlap of the initial state $\ket{0}$ and final state $\bra{0}$ with the projector from the middle region: $2\inner{0}{+}\inner{+}{0} = 1$. 

We may analyze the $\bx_1$ variables similarly.
Because the long region where $\bx_1 \approx \bx_2$ again gives a projector $\ket{+}\bra{+}$, and because the transient regions satisfy the same relations between the variables as the transition regions from the $\bx_2$ analysis, the same arguments we made about the transition regions for $\bx_2$ goes through for the transient behaviors of $\bx_1$, and the Pfaffian does not depend on the precise form of the transient behaviors.
The remaining Hamiltonian contribution from $X$ again cancels the bulk term in the transient regions.
There are no initial or final fixed point regions for $\bx_1$, so the total contribution is from another projector overlap with the relevant initial and final states: $2\inner{+}{+}\inner{+}{+} = 2$.
It may seem redundant to analyze the $\bx_1$ variables separately as we have done here, since the $A_1$ Pfaffian term involves the total $\bx$ variables like the $A_2$ Pfaffian.
However, it was important here to conclude that the transient behaviors do not contribute any time-dependence at $O(\lambda)$, and we actually obtained a constant result that is independent of $T$.

Thus, again using the normalization \eqref{eq:normalization}, from the connected solution we have a contribution 
\begin{equation}
    \Tr \sigma_L^2\Big|_{\text{conn}} = 2^{-N(1-\lambda)} .
\end{equation}

\subsection{Other tunneling solutions}\label{sec:lattice-solns}

\begin{figure}
    \centering
    \includegraphics[width=.49\textwidth]{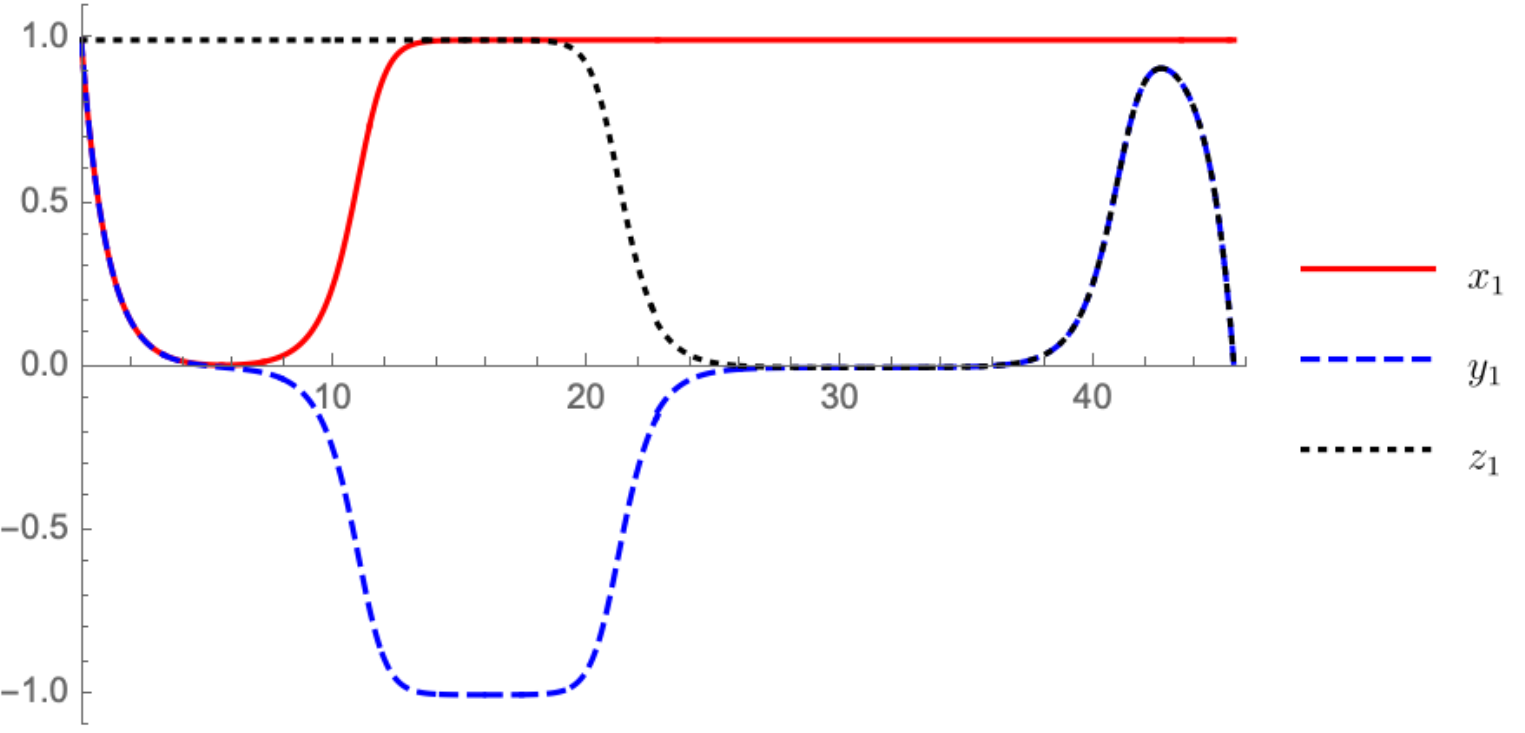}
    \includegraphics[width=.49\textwidth]{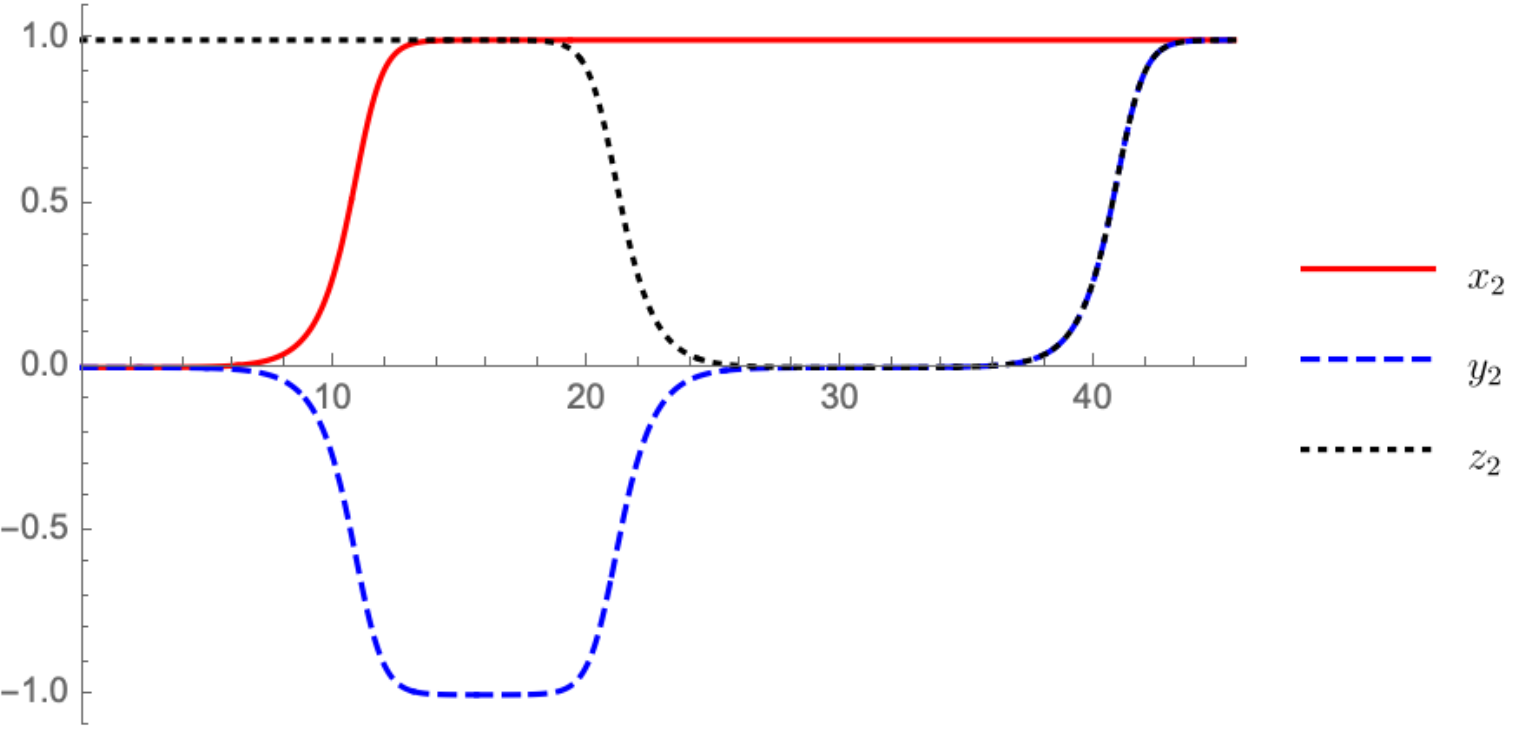}
    \caption{\small{The DW solution for the $\bx_1$ (left) and $\bx_2$ (right) variables with $J=q=4$, $r = 1-10^{-8}$, $\lambda = 10^{-4}$, and $T \approx 45.5$. There are two long regions with $\bx_2 \approx (0,0,1)$ and $(1,0,0)$, although the $(0,0,1)$ region is a little smaller for this value of $\lambda$.  The initial transient behavior for $\bx_1$ matches the disconnected solution in Figure~\ref{fig:disk-soln} while the final transient matches the connected solution in Figure~\ref{fig:connected-soln}.}}
    \label{fig:disk-wh-soln}
\end{figure}

\begin{figure}
    \centering
    \includegraphics[width=.49\textwidth]{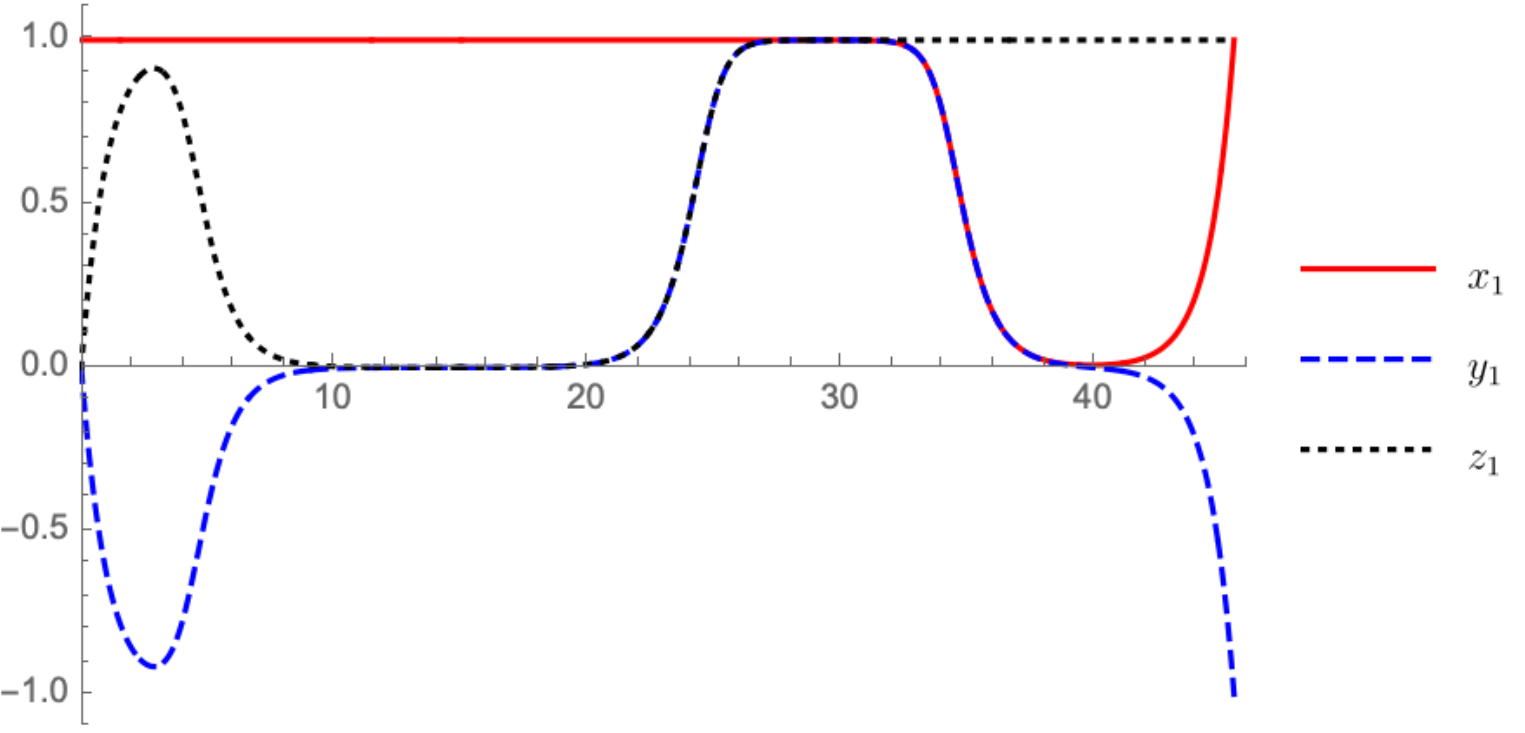}
    \includegraphics[width=.49\textwidth]{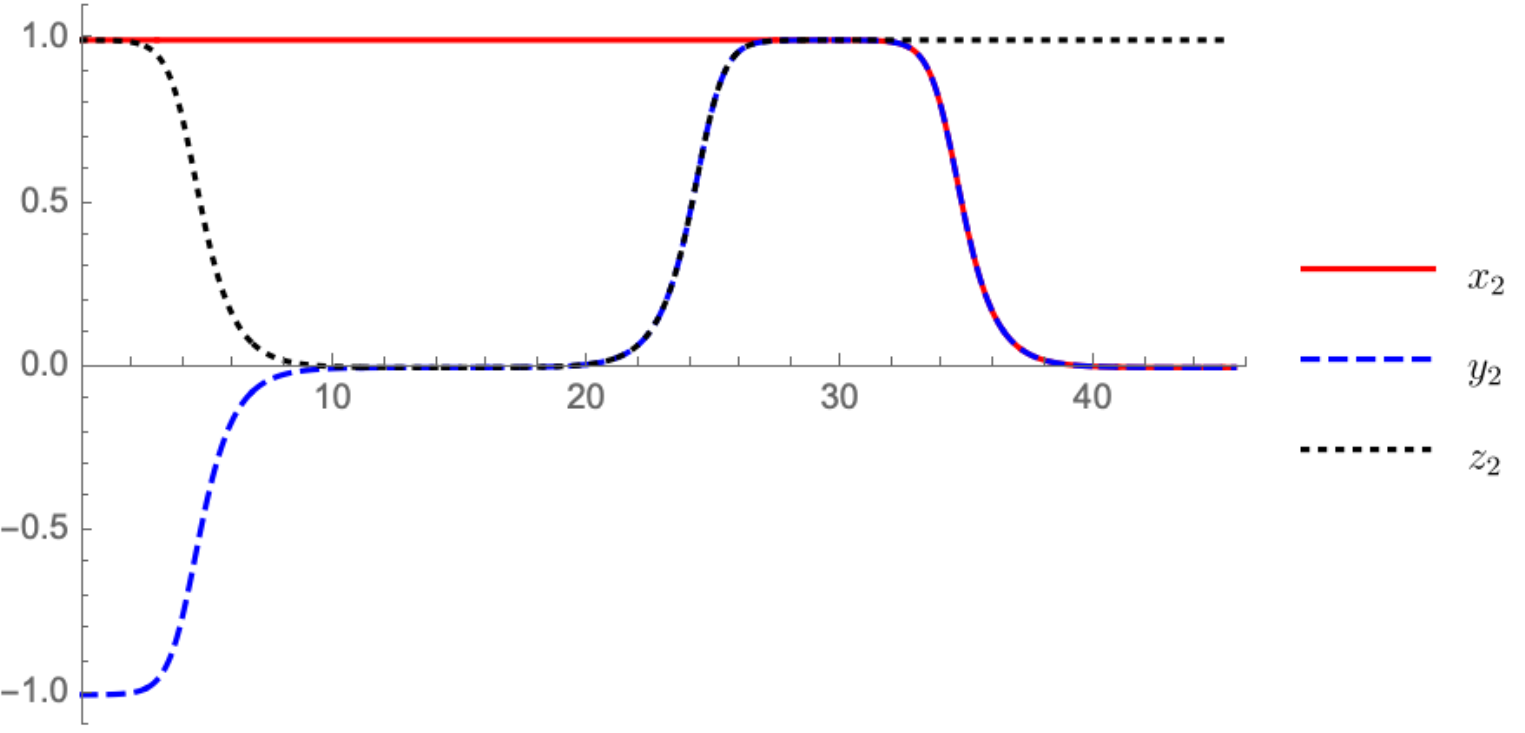}
    \caption{\small{The WD solution for the $\bx_1$ (left) and $\bx_2$ (right) variables with $J=q=4$, $r = 1-10^{-8}$, $\lambda = 10^{-4}$, and $T \approx 45.5$. There are two long regions with $\bx_2 \approx (1,0,0)$ and $(0,0,1)$, though the $(0,0,1)$ region is a little smaller for this value of $\lambda$.  The initial transient behavior for $\bx_1$ matches the connected solution in Figure~\ref{fig:connected-soln} while the final transient matches the disconnected solution in Figure~\ref{fig:disk-soln}.}}
    \label{fig:wh-disk-soln}
\end{figure}

There are also solutions which tunnel from $\bx_2 \approx (0,0,1)$ to $\bx_2 \approx (1,1,1)$ and from $\bx_2 \approx (1,-1,1)$ to $(0,0,1)$.
We name these the ``DW'' and ``WD'' solutions, respectively, after the order of transient behavior which occurs for the $\bx_1$ variable: the first has ``Disk'' initial transient behavior and ``Wormhole'' final transient behavior, while the second has the opposite ordering.
The DW solution is shown in Figure~\ref{fig:disk-wh-soln} while the WD solution is shown in Figure~\ref{fig:wh-disk-soln}.
The contribution of these solutions to $\Tr \sigma_L^2$ can be evaluated in the same approximate manner as Section~\ref{sec:wormhole-soln}.

We begin with the DW solution in Figure~\ref{fig:disk-wh-soln}.
The $\bx_2$ variables has the same long region with $\bx_2 \approx (1,0,0)$ which appears in the connected solution (Figure~\ref{fig:connected-soln}), and by the same reasoning as in Section~\ref{sec:wormhole-soln} we conclude that this region yields for the path integral the projector $\ket{+}\bra{+}$.   
Similarly, the long region with $\bx_2 \approx (0,0,1)$ gives a projector $\ket{0}\bra{0}$.
These two projectors cancel the transition regions and the other constant regions associated with other fixed points, and we get the overlap $2\inner{0}{0}\inner{0}{+}\inner{+}{0} = 1$.
The $\bx_1$ variables have a leading contribution determined by simply changing the initial and final states: $2\inner{+}{0}\inner{0}{+}\inner{+}{+} = 1$.
Thus, including the normalization \eqref{eq:normalization}, we have the additional suppression $2^{-N}$ for the DW solution:\footnote{We are neglecting the one-loop determinant here.  As shown in \cite{Stanford:2021bhl}, this determinant can lead to an overall minus sign for some of these subleading solutions.  Because they are subleading anyway, we will omit this effect, which does not affect the disconnected or connected saddle points.}
\begin{equation}
    \Tr \sigma_L^2 \Big|_{\text{DW}} = 2^{-N}.
\end{equation}
We will not bother to compute the $O(\lambda)$ contribution from the transient behaviors of $\bx_1$ in the $A_1$ Pfaffian (which could lead to nontrivial $T$ dependence), since this solution is already highly suppressed compared to the connected one in Section~\ref{sec:wormhole-soln}.

\begin{figure}
    \centering
    \includegraphics[width=.8\textwidth]{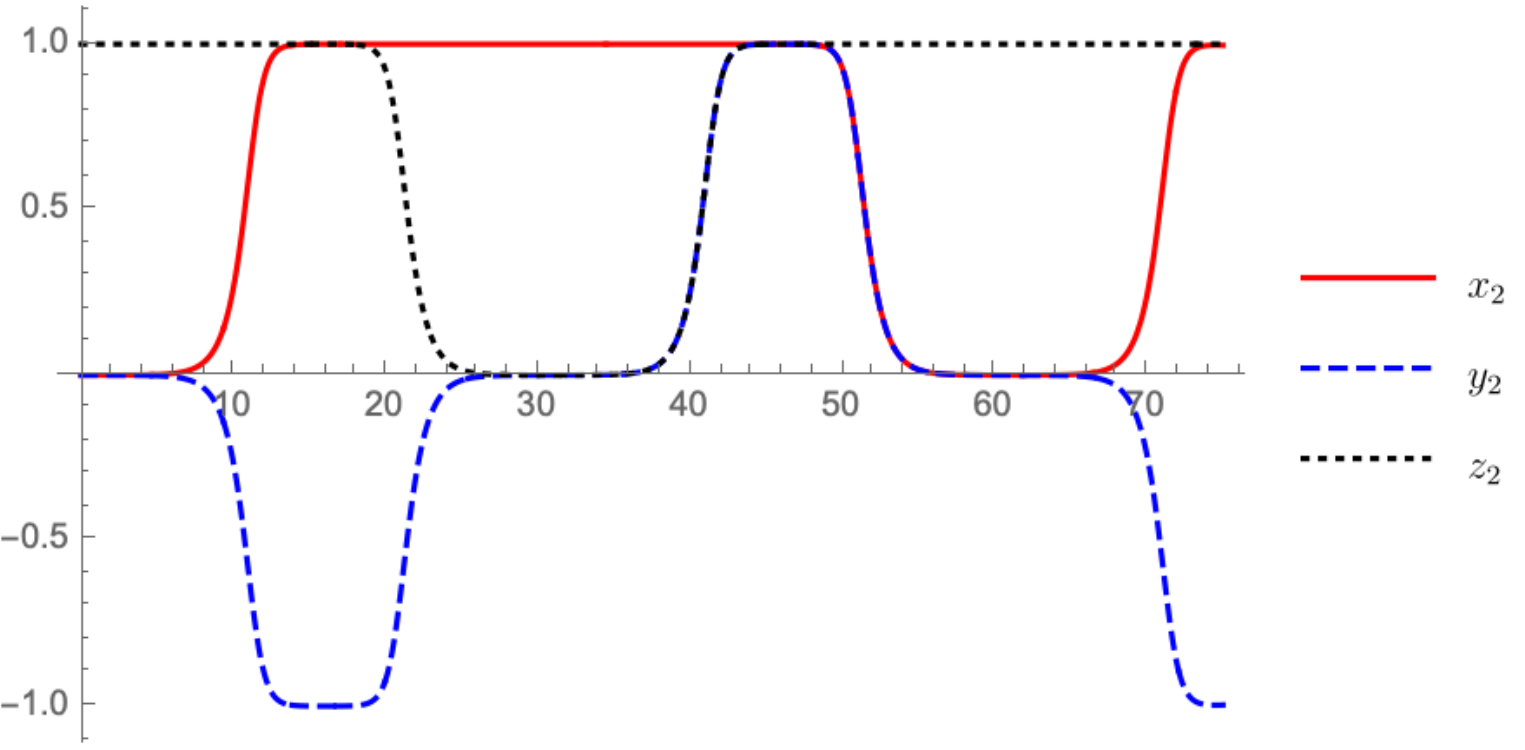}
    \caption{\small{A full period of the $\bx_2$ variables with $J=q=4$, $r=1-10^{-8}$, $\lambda = 10^{-4}$.
    This periodic segment can be inserted into the disconnected, connected, DW, or WD solutions however many times we like to produce new solutions (for different values of $T$) that are suppressed with powers of $2^{-kN}$, where $k$ is the number of inserted periods.
    }}
    \label{fig:period}
\end{figure}

The WD solution can be analyzed similarly and also has a $2^{-N}$ leading suppression.
Thus, both the DW and WD solutions are subleading compared to the connected solution from Section~\ref{sec:wormhole-soln}.
There are also even more highly suppressed solutions which can be formed by inserting additional periods into any of the four solutions we have discussed up to this point.
A full period of the $\bx_2$ variables is shown in Figure~\ref{fig:period}.
By the same approximate reasoning, inserting a full period in the solution will suppress the contribution to the path integral by an additional $2^{-N}$.

Interestingly, the long regions of the solution have a minimum length which scales like the scrambling time $T_s \sim (1/J)\log N$.
What this means is that they are actually not solutions for all values of $T$.
For instance, the connected solution in Section~\ref{sec:wormhole-soln} is only a solution for $T > (1/J)\log N$.
A configuration with $k$ long regions will not appear as a solution until $T > (k/J) \log N$.
This lattice of critical times is interesting from a unitarity perspective.
These subleading saddles are necessary to ensure the total Brownian evolution is unitary, so an experimentalist with access to only on-shell configurations will discover that it is impossible to verify unitarity with accuracy better than $2^{-kN}$ until at least $T > (k/J) \log N$.

\subsection{Summary}

\begin{figure}[t]
    \centering
    \includegraphics[width=.7\textwidth]{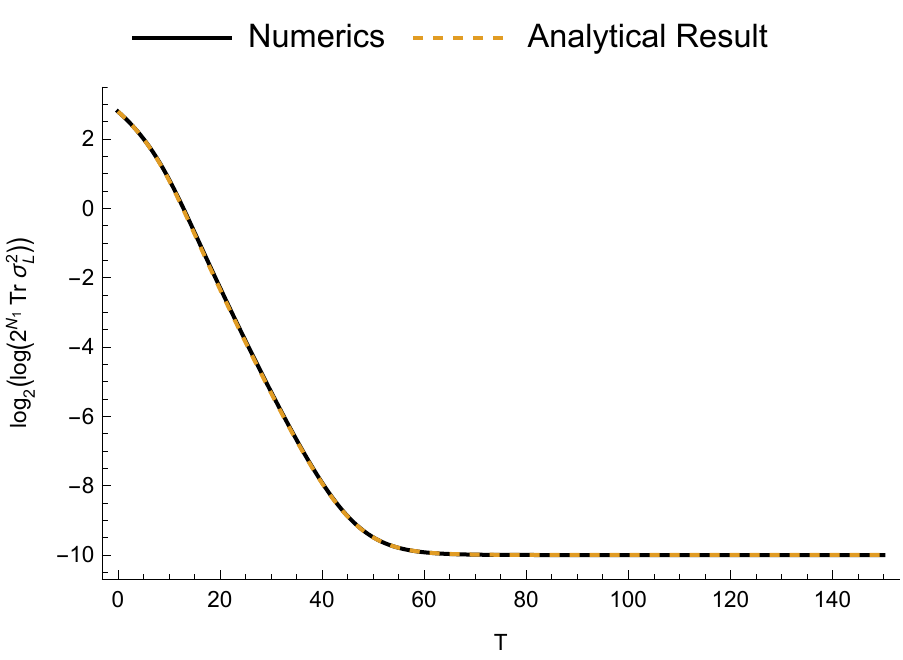}\caption{\small{A comparison between numerical evaluation of $\log_2[\log (2^{N1} \Tr \sigma_L^2)]$ with the approximate analytical result of equation  (\ref{eq:Trsigmal}) for $N = 30$ and $\lambda = 1/3$. At late times, the quantity saturates to $-(N_2 - N_1) = -(1-2\lambda)N = -10$.}}
    \label{fig:comparison}
\end{figure}

We have shown that the leading $T$ dependence of the purity which controls the mutual purity in Brownian SYK is\footnote{In this analysis, we have purposefully ignored the presence of discrete symmetries.  The presence of such symmetries generically prevents the time evolution from covering the entire unitary group. Following \cite{Stanford:2021bhl}, we can adapt the analysis of Brownian SYK so that the time evolution covers the entire unitary group by only including the saddle points we have discussed.  Incorporating the discrete symmetries of the SYK model requires additional saddle points \cite{Stanford:2021bhl}.  This means our results are effectively valid for an SYK-like model with no discrete symmetries which does end up covering the whole unitary group.}
\begin{equation}\label{eq:Trsigmal}
    \Tr \sigma_L^2 = \left( \frac{1 + e^{-JT/2^{q-1}}}{2} \right)^{\lambda N} + \frac{\Theta(JT-\log N)}{2^{(1-\lambda)N}} + \dots ,
\end{equation}
where the first term comes from the disconnected saddle point, the second term from the leading connected saddle point, and the dots represent further subleading solutions that are suppressed in powers of $2^{-N}$.
We present a comparison of this saddle point analysis with an exact numerical computation of $\Tr \sigma_L^2$ in Figure~\ref{fig:comparison}.
The form of \eqref{eq:Trsigmal} means $\cF_{\Psi'}(\re : \en)$ is initially $O(1)$ and subsequently decays for a polynomial $T \sim N/J$ amount of time.
When $T > N/J$, the connected solution begins to dominate and leads to an exponentially small mutual purity.\footnote{Recall that that $\Tr \sigma_L^2$ enters in $\cF_{\Psi'}(\re : \en)$ along with a subtraction of a baseline value, and so the contribution which dominates $\cF_{\Psi'}(\re: \en)$ is not necessarily the one which makes the largest contribution to $\Tr \sigma_L^2$.}
Thus, when the encoding complexity is sufficiently large, the code is robustly protected from the erasure of $A_1$.  At multiples of the scrambling time $T_s \sim (1/J)\log N$, subleading contributions become genuine on-shell solutions of the equations of motion, though these contributions never dominate $\cF_{\Psi'}(\re : \en)$.

\section{Discussion}\label{sec:disc}

We have studied the error correction properties of Brownian SYK quantum codes against the erasure of a small number of qubits, but we expect our results to be valid more generally for generic, low-rank errors with no prior access to the encoding map. As a measure of quantum error correction, we computed the mutual purity $\cF_{\Psi'}(\re : \en)$, which is related to the purity $\Tr \sigma_L^2$, where $\sigma = VV^\dagger / d_\co$  is the density matrix built from the encoding map $V$, $\sigma_L = \Tr_R \sigma$, and $R$ is a small fraction of the physical Hilbert space which is being erased.
In codes defined using Brownian SYK time evolution, which have a linearly growing encoding complexity  ---  mimicking the expected behavior of the bulk-to-boundary map for an infalling observer in AdS/CFT --- this purity is related to a four-contour Lorentzian (Schwinger-Keldysh) path integral.
We found two special saddle point solutions to the large $N$ equations of motion in Brownian SYK --- analogous to the disconnected disks and connected wormhole geometries in JT gravity --- which dominate this path integral.
At early times $T \ll N/J$, the disconnected solution gives an exponentially decaying value for the mutual purity, while at late times the connected solution dominates and gives a constant, exponentially small mutual purity. Thus, when the encoding complexity is sufficiently large, we find emergent, ``complexity-protected'' quantum error correction against generic, low-rank errors with no prior access to the encoding map. We should emphasize that it is important that the error does not have access to the encoding map -- with prior access, it is possible to violate the above conclusions. 

\subsection{Relation to previous work}

Understanding how the complexity of an encoding operator affects certain error correction properties of the code is a problem that has been explored previously from a variety of viewpoints.
The most common method of studying codes with increasing complexity is to employ the randomization trick as we have done, where one instead considers a one-parameter family of ensembles of codes with increasing complexity and studies ensemble-averaged properties.

For instance, \cite{Brown:2013cmp,Brown:2013} argued that $n$-qubit random quantum circuits with $O(n \log^2 n)$ two-qubit gates and $O(\log^3 n)$ depth can encode $k$ qubits into $n$ while correcting erasure errors on $d$ qubits where
\begin{equation}
    \frac{k}{n} < 1 - \frac{d}{n} \log_2 3 - h(d/n) ,
\label{eq:distance-bound}
\end{equation}
with $h(x)$ being the binary entropy function
\begin{equation}
    h(x) \equiv -x\log x - (1-x)\log (1-x) \,  . 
\end{equation}
In our analysis, we  studied a random error with $k/n = d/n = \lambda$, and with these replacements the inequality \eqref{eq:distance-bound} is true for roughly $\lambda \leq 1/5$.
It would be interesting to understand whether our analytics can be extended to this rather large value of $\lambda$ without the need for novel techniques, although the numerical results in Figure~\ref{fig:comparison} suggest we may have an accurate picture of the Brownian theory even when $\lambda = 1/3$.
At any rate, it appears that the Brownian codes we have studied in this work are able to \emph{approximately} (with error of order $1/n$) correct errors on a fraction $\lambda$ of the physical qubits with a depth $T \sim (1/J) \log n$. This polynomial improvement in depth, if true, is likely due to differences in how the random two-qubit quantum circuit theory of \cite{Brown:2013} and the Brownian SYK theory scramble quantum information.

More recently, \cite{Gullans:2021} studied low depth random circuits with spatial connectivity restrictions in various spatial dimensions $D$ as stabilizer codes.
They discovered that such circuits can correct fairly large erasure errors (converging to both the optimal threshold and zero failure probability at large $n$) with a depth of just $O(\log n)$ for $D \geq 2$.
These results are  similar to ours, although we have no restriction on spatial connectivity, but rather a restriction on the number of fermions which can couple in the Hamiltonian.
It would be interesting to understand if there is a relation between the universality for $D \geq 2$ found in \cite{Gullans:2021} and the expected universality of our results for $q \geq 4$.
A significant difference of our analysis compared with \cite{Gullans:2021} is that we do not restrict ourselves to stabilizer codes, though we also have not studied the decoding problem in any detail.

Beyond questions of depth, we may also consider the total gate complexity of efficient quantum codes.
Several bounds on this complexity exist for stabilizer codes \cite{Cleve:1997eff,Aaronson:2004imp,Kuo:2019enc} and their generalizations \cite{Brun:2006cor,Kuo:2019enc}.
In particular, for a generic stabilizer code encoding $k$ qubits into $n$, \cite{Kuo:2019enc} showed that $O(n(n-k)/\log n)$ gates are sufficient.
Entanglement-assisted stabilizer codes were also studied in \cite{Kuo:2019enc} and were shown to have gate complexity linear in the number of additional entangled qubits $c$, with $O(n(n-k+c)/\log n)$ gates.
As we have not restricted ourselves to stabilizer circuits, our gate complexity is not expected to have such small polynomial asymptotic behavior.\footnote{Efficient Hamiltonian simulation of Brownian SYK would likely involve discretization of the contact correlation $\delta(t-t')$ in the variance, along with a sparse query model like the one studied in \cite{Berry:2015ham,Berry:2019tim}.  Because the sparsity of the full SYK Hamiltonian scales with $N^q$, we do not expect simulation to be efficient compared to stabilizer circuits.}
However, if we used a sparse SYK model instead \cite{Xu:2020shn}, we may achieve equal or better gate complexity compared to stabilizer circuits.  This issue deserves further study as it would represent an interesting development in efficient random code design.

Our work is also closely related to  measurement-induced phase transitions which have recently been studied extensively in the condensed matter community (see for instance \cite{Choi:2020prl, Gullans:2019zdf}). In these studies, the quantum circuit usually consists of local unitary gates with some quenched disorder and forms a brickwall pattern. These local unitaries are interspersed with local measurements, which are viewed as ``errors''. The long range entanglement generated by the random unitary gates is identified as the volume-law phase, suitable for quantum error correction. However, a transition to a short-range entanglement phase can occur when measurement rate is high, i.e. when the error rate is high enough to disentangle different subsystems. The volume-law to area-law transition is identified as a transition in quantum error correction, when the error rate exceeds a critical value \cite{Choi:2020prl, Gullans:2019zdf}. In essence, the size of the Hilbert space of the principal quantum system needs to be large enough (spatial) and the time for the unitary gates need be long enough (temporal) to scramble the information so that the entanglement is robust against local disturbances. It would be interesting to compare these results with those presented here.

In another direction, ensembles of encoding maps that satisfy some global symmetry have also been explored \cite{Eastin:2009,Faist:2019ahr}.
The general idea is that there is a tension between the existence of a continuous symmetry leaving the encoding map invariant and strong protection against erasure errors.
However, approximate error correction can be achieved in certain circumstances \cite{Kong:2021wau}.
In Brownian SYK, there are discrete global symmetries (which we did not include in the analysis since we were interested in covering the entire unitary group) but no continuous symmetries, allowing us to avoid these no-go arguments. 
However, it is easy to implement continuous symmetries in analogues of the SYK model; for instance, SYK with complex fermions satisfies a $U(1)$ global symmetry \cite{Gu:2019jub}.
It would be interesting to understand the error correction behavior of a complex analogue of Brownian SYK to further elucidate the tension between codes with continuous symmetries and erasure error correction.

\subsection{Pseudorandom codes}

Our results seem to suggest that after a polynomial time, a random quantum circuit, which likely has polynomial circuit complexity, has powerful error correction properties that are  essentially as good as a Haar random unitary code, which likely has exponential complexity.
One explanation for why this is possible may be that the majority of unitary operators with polynomial complexity are in fact pseudorandom unitary operators, and a simple test of error correction properties cannot distinguish polynomially complex pseudorandom unitary operators from  unitary operators of exponential complexity.

A pseudorandom unitary operator is, roughly speaking, an operator which has polynomial complexity but which cannot be distinguished from one with exponential complexity by any sort of simple test which can be implemented efficiently.
The transition between disconnected and connected solutions that we found, hints at a sharp transition point where most random circuits with complexity less than some polynomial ($N$, for the purity transition) are not pseudorandom, while the typical circuit and perhaps the majority of circuits above that critical complexity are in fact pseudorandom, at least for the purposes of error correction.  
It would be very interesting to understand in more detail what properties of Haar random circuits can be reproduced by such low complexity Brownian circuits.

\subsection{Complexity and the geometry of the entanglement wedge}

From the AdS/CFT point of view, it would be very interesting to understand the bulk significance of our results; indeed, one of our main motivations in this paper was to understand the geometry of Figure~\ref{fig:PL1} in terms of quantum error correction. Following \cite{Kim:2020cds, Balasubramanian:2022fiy}, we expect that this error correction is a sign of ``causal inaccessibility'' from the boundary subregion. By this, we mean that including backreaction from turning on simple sources in the asymptotic boundary does not render the relevant degrees of freedom causally accessible from the boundary; the mechanism behind this is that the relevant bulk degrees of freedom lie behind a non-minimal quantum extremal surface. In our calculation, we encountered two significant complexity scales, i.e., the mutual purity becomes $O(1/N)$ at $T \sim \log N$, and the mutual purity saturates to an exponentially small plateau at a much larger time-scale. It is tempting to speculate that these thresholds have natural bulk interpretations: the $\log N$ time-scale could correspond to the bulk degrees of freedom crossing the causal horizon, while the plateau could correspond to the bulk degrees of freedom crossing over to the python's lunch. In a similar vein, the lattice of subleading solutions we found may also have a geometric meaning, although it is less clear because they do not dominate the calculation of the crucial quantity $\cF_{\Psi'}(\re : \en)$.

\subsection*{Acknowledgments}

We thank Bartek Czech, Abhijit Gadde, Issac Kim, Gautam Mandal, Shiraz Minwalla, Pranab Sen and Sandip Trivedi for helpful discussions.
VB is supported in part by the Department of Energy through grant DE-SC0013528 and grant QuantISED DE-SC0020360, as well as the Simons Foundation through the It From Qubit Collaboration (Grant No. 38559). 
AK is supported by the Simons Foundation through the It from Qubit Collaboration.
CL is supported by the Department of Energy through QuantISED grant DE-SC0020360. OP and HR are supported by the Department of Atomic Energy, Government of India, under project
identification number RTI 4002.

\section*{} 

\appendix
\section{Hamiltonian formalism in Brownian SYK}\label{sec:Hamiltonian}
In this Appendix, we will reproduce some of the results of Section~\ref{sec:typical-brownian} from a different point of view. Recall that after averaging over the couplings $J_{a_1 \dots a_q}(t)$ in the Brownian SYK model, we obtain the following effective action:
\beq
I = \frac{1}{2} \int_0^T \d t \, \psi^{(j)}_a \partial_t  \psi^{(j)}_a - \frac{i^qJ(q-1)!}{2N^{q-1}} \int_0^T \d t   s_j s_{j'} \,\psi^{(j)}_{a_1\cdots a_q}(t) \psi^{(j')}_{a_1\cdots a_q}(t).
\eeq
From the action, we can read off an ``effective Hamiltonian'':
\begin{equation}
H_{\text{eff}} = -i^q \frac{J(q-1)!}{2N^{q-1}} \, \sum_{j,k} \sum_{a_1 < a_2 \dots < a_q } s_j s_k \, \psi^{(j)}_{a_1\dots a_q} \psi^{(k)}_{a_1\dots a_q}, \label{eq:typhamiltonian}
\end{equation}
where, $\psi^{(j)}_{a_1\dots a_q} = \prod_{i = 1}^{q}\psi_{a_i}^{(j)}$. We note that $H_{\text{eff}}$ commutes with the fermion parity operator defined in equation (\ref{parity1}) and (\ref{parity2}). Therefore, we can write $H_{\text{eff}}$ in terms of the Pauli matrices defined as follows:
\begin{equation}
		\psi^{(1)}_a \psi^{(2)}_a = -\frac{i}{2} X_a, \quad \psi^{(1)}_a \psi^{(3)}_a = \frac{i}{2} Y_a, \quad \psi^{(1)}_a \psi^{(4)}_a = -\frac{i}{2} Z_a.
\end{equation}
After substituting the above relations in (\ref{eq:typhamiltonian}), the effective Hamiltonian can be written as
\begin{equation}
	H_{\text{eff}} = -\frac{J(q-1)!}{(2N)^{q-1}} \sum_{a_1 < a_2 \dots < a_q}  \left( X_{a_1} \dots X_{a_q} - i^q Y_{a_1} \dots Y_{a_q}  + Z_{a_1} \dots Z_{a_q} - 1 \right) ,
\end{equation}
where the last term (proportional to the identity) comes from the $j=k$ terms. $\Tr \sigma_L^2$ can now be written as the Euclidean transition amplitude:
\begin{equation}
\Tr \sigma_L^2 = \bra{\psi} \exp(-H_{\text{eff}}T) \ket{\psi} , \label{eq:trsigma0}
\end{equation}
where the initial and final states are dictated by the boundary conditions in the path integral, and are given by 
\beq
\ket{\psi} = \ket{+}^{N_1} \ket{0}^{N_2}.
\eeq

\subsection{Disconnected and connected solutions}
In the above expression, we can separate contributions from the ground states and the excited states as follows:
\begin{equation}
	\begin{split}
		\Tr \sigma_L^2 =  \sum_{n} |\inner{\psi}{g_n}|^2 + \sum_{k} \exp{(-E_k T)} |\inner{\psi}{e_k}|^2 . 
	\end{split}\label{eq:trsigma1}
\end{equation}
Here, $g_n$ are all the ground states of $H_{\text{eff}}$ (which all have zero energy) and $e_k$ are the excited states with energies $E_k$. We first look at the contribution from the ground states. The set of ground states depends on whether we choose $q = 4k$ or $q = 4k + 2$ but the following two ground states contribute to the leading order independently of $q$: 
\begin{equation}
\begin{split}
\ket{g_1} &= \ket{0}^N , \\
\ket{g_2} &= \ket{+}^N .
\end{split}
\end{equation} 
Therefore, the contribution from the ground states is
\begin{equation}
\begin{split}
\sum_{n} |\inner{\psi}{g_n}|^2 &\approx \inner{\psi}{g_1}|^2 +  |\inner{\psi}{g_2}|^2 \\ & =\frac{1}{2^{N_1}} + \frac{1}{2^{N_2}},
\end{split}
\end{equation} 
which reproduces the two leading order terms (i.e., the disconnected and the connected contributions) in the Haar ensemble. The contribution from excitations near the ground states can be approximated in the following manner. The Hamiltonian can be written in terms of the ladder operator as
\begin{equation}
\begin{aligned} 
		H &= -\frac{2JN}{q} \left[\left(\frac{S_x}{N}\right)^q - \left(\frac{iS_y}{N}\right)^q  + \left(\frac{S_z}{N}\right)^q - \frac{1}{2^q}\right] + O\left(\frac{1}{N}\right) , \label{eq:Hamiltonianap}
		\end{aligned}
\end{equation}
where $S_x = \sum_a X_a/2$, $S_y = \sum_a Y_a/2$, and $S_z = \sum_a Z_a/2$. The matrix elements of the first two terms in the Hamiltonian with excited states near $\ket{g_1}$ are suppressed by a factor of $ 1/N^{q/2}$ and can be ignored for $q \geq 4$ at leading order in $1/N$. Thus, the Hamiltonian up to $O(1/N)$ corrections is
\begin{equation}
		H = -\frac{2JN}{q} \left[ \left(\frac{S_z}{N}\right)^q - \frac{1}{2^q}\right] . \label{eq:Hamiltonian1byN}
\end{equation} 
The contribution from the states near $\ket{g_1}$ is:
\begin{equation}
\begin{split}
\Tr_{g_1} \sigma_L^2 &= 	 \sum_{k} \exp{(-E_k T)} |\inner{\psi}{e^{g_1}_k}|^2 \\ 
	 &\approx \sum_{k=0}^{N_1} \frac{1}{2^{N_1}} \binom{N_1}{k} \exp\left( -\frac{JkT}{2^{q-2}}\right) \\
	 &= \left( \frac{1 + \exp \left( -\frac{JT}{2^{q-2}}\right)}{2} \right)^{N_1} ,
\end{split} \label{eq:tracesaddle1}
\end{equation}
where $\ket{e^{g_1}_k}$ denotes the $k^{th}$ excited state near $\ket{g_1}$, explicitly given by a choice of $k$ qubits which are flipped to $\ket{1}$ from $\ket{0}$.
These $k$ must come from the first $N_1$ qubits to give a nonvanishing overlap $\inner{\psi}{e_k^{g_1}}$.

Similarly, we can compute the correction due to the excited states $\ket{e^{g_2}_k}$ near $\ket{g_2}$. 
The perturbative Hamiltonian is now
\begin{equation}
    H = -\frac{2JN}{q} \left[ \left( \frac{S_x}{N} \right)^q - \frac{1}{2^q} \right]\ ,
\label{eq:g2-hamiltonian}
\end{equation}
and the excited states are formed by flipping $k$ qubits to $\ket{-}$ from $\ket{+}$, where these must come from the last $N_2$ qubits to give a nonvanishing overlap with the boundary state. 
We get
\begin{equation}
\begin{split}
		\Tr_{g_2} \sigma_L^2 &= \sum_{k} \exp{(-E_k T)} |\inner{\psi}{e^{g_2}_k}|^2 \\ &\approx  \frac{1}{2^{N_2}} \left( 1 + N_2 \exp\left(-\frac{JT}{2^{q-2}} \right)+ \dots \right) .
\end{split}
\label{eq:g2-contribution}
\end{equation}
Note that the contribution from the second and higher excited states (denoted here by ellipsis) is not negligible. Moreover, unlike the case of the disconnected saddle where $N_1/N = \lambda$ was a small parameter, we cannot resum all the contributions from higher excited states near $\ket{g_2}$. Since $N_2/N \sim 1$, one must also take the quantum corrections into account.  Nevertheless, the above expression is sufficient to infer that the ground state contribution dominates when $T > \frac{2^{q-2}}{J} \log(N_2) $. 

Summing up these contributions we have the  following result for $\Tr \sigma_L^2$:
\begin{equation}
\begin{split}
		\Tr \sigma_L^2 &\approx \Tr_{g_1} \sigma_L^2 + \Tr_{g_2} \sigma_L^2 \\ & \approx \left( \frac{1 + \exp \left( -\frac{JT}{2^{q-2}}\right)}{2} \right)^{N_1} +\, \frac{1}{2^{N_2}} \left( 1 + N_2 \exp\left(-\frac{JT}{2^{q-2}} \right)+ \dots \right) .
\end{split}\label{eq:trace}
\end{equation}
The $T$ dependent term proportional to $N_2/2^{N_2}$ is a contribution from corrections to the projector approximation to the long region we made in Section~\ref{sec:wormhole-soln}.
We could have incorporated such terms in the path integral saddle point approximation of Section~\ref{sec:wormhole-soln} by writing the long region as a projector $\ket{+}\bra{+}$ plus an exponentially suppressed correction $e^{-JT} \ket{-}\bra{-}$.
However, the evaluation of the saddle point including this correction is difficult because the transient regions no longer cancel against the $\ket{-}\bra{-}$ operator, so this term induces large corrections which depend sensitively on the transient shape.
This is the path integral analogue of the Hamiltonian picture difficulty we described under \eqref{eq:g2-contribution}.
Of course, these corrections are only important before $T < (1/J)\log N$, when the connected configuration is not actually a solution of the equations.
By the time the connected configuration becomes a genuine saddle point, this $T$ dependence is subleading and the constant $2^{-N_2}$ term dominates up to possible $O(1)$ factors just as $T$ crosses $(1/J) \log N$.
Furthermore, by the time the connected solution actually dominates the mutual purity, these terms are suppressed by an even stronger factor of $e^{-N}$ compared to the constant $2^{-N_2}$ term.

\subsection{One-loop determinant around disconnected solution}

In the Hamiltonian picture, the one-loop determinant is related to corrections in the energy eigenstates and the corresponding eigenvalues near the ground state $\ket{g_1}$. From equation (\ref{eq:Hamiltonianap}), we see that the energy eigenstates $\ket{e^{g_1}_k}$ gain corrections from the first two terms related to the ladder operators. However, since they are suppressed by a factor of $1/\sqrt{N^{q}}$ we can ignore these corrections. The correction to energy eigenvalues can be computed by expanding the $S_z^q$ term to $O(1/N^2)$:
\begin{equation}
	E_k = \frac{J}{2^{q-2}} \left(k  - (q-1) \frac{k^2}{N} \right) + O(N^{-2}) .
\end{equation} 
Thus, the contribution to $\Tr \sigma_L^2$ from the first saddle including corrections at $O(1/N)$ is 
\begin{equation}
	\Tr^{(1)}_{g_1} \sigma_L^2  = \sum_{k= 1}^{N_1} \binom{N_1}{k} \exp  \left(-\frac{JTN_1}{2^{q-2}} \left(\frac{k}{N_1}  - (q-1)\lambda \frac{k^2}{N_1^2} \right)  \right) .
\end{equation}
To extract the one-loop determinant from the above expression we divide it by the classical saddle point result in equation (\ref{eq:tracesaddle1}) and take the large $N_1$ limit keeping $\lambda = \frac{N_1}{N}$ fixed. Define $F(T)$ as
\begin{equation}
\begin{split}
	F(T) &\equiv \frac{\Tr^{(1)}_{g_1}\sigma_L^2}{\Tr^{(0)}_{g_1}\sigma_L^2} \\ 
	&= \frac{\sum_{k= 1}^{N_1} \binom{N_1}{k} \exp  \left(-\frac{JTN_1}{2^{q-2}} \left(\frac{k}{N_1}  - (q-1)\lambda \frac{k^2}{N_1^2} \right)  \right)}{\sum_{m= 1}^{N_1} \binom{N_1}{m} \exp  \left(-\frac{JN_1T}{2^{q-2}}\frac{m}{N_1} \right)} .
\end{split}
\end{equation}
In the large $N_1$ limit, we use Stirling's approximation for the factorial terms and replace the sum over $k$ by an integral over $x \equiv k/N_1$ to write $F(T)$ as follows:
\begin{equation}\begin{split}
	F(T) &\approx \frac{\int_0^1 \d x \frac{1}{\sqrt{x(1-x)}} \exp\left[-N_1 f(x) \right]}{\int_0^1 \d x \frac{1}{\sqrt{x(1-x)}} \exp\left[-N_1 g(x) \right]} , \\ 
\end{split}
\end{equation}
where we have defined the functions
\begin{equation}
\begin{split}
	g(x) &= x \log x + (1 -x) \log (1-x) + \frac{JT}{2^{q-2}}x, \\
	f(x) &= g(x) - \lambda \frac{JT(q-1)}{2^{q-2}}x^2 .
 \end{split}
\end{equation}
The integrals can be evaluated in the saddle point approximation and we get the following result:
\begin{equation}
	F(T) \approx \sqrt{\frac{g''(x_g) x_g(1-x_g)}{f''(x_f)x_f(1-x_f)}} \exp\left[-N_1\left( f(x_f) - g(x_g) \right)  \right] .
\end{equation}
Here, $x_f$ and $x_g$ are saddle points of $f(x)$ and $g(x)$ respectively. Since $f(x)$ and $g(x)$ differ by a term proportional to $\lambda$, we can evaluate $F(T)$ perturbatively in $\lambda$. We have the following equations:
\begin{equation}
\begin{split}
	g'(x_g) = 0 &\implies \frac{x_g}{1 - x_g} = \exp\left(-\frac{JT}{2^{q-2}}\right), \\
	f'(x_f) = 0 &\implies g'(x_f) = \lambda \frac{JT(q-1)}{2^{q-3}}x_f \\
	&\implies (x_f - x_g)g''(x_g) = \lambda \frac{JT(q-1)}{2^{q-3}}x_g + O(\lambda^2) .
\end{split}
\end{equation}
Another useful relation is 
\begin{equation}
	g'''(x) = -g''(x) h(x)  ,
\end{equation}
where the function $h(x)$ is
\begin{equation}
h(x) = \frac{1}{x} - \frac{1}{1-x} .
\end{equation}
Using the above relations, we first evaluate the term in the square root. 
\begin{equation}
	\begin{split}
		 \frac{f''(x_f) x_f(1-x_f)}{g''(x_g)x_g(1-x_g)} &= 1 + \frac{g'''(x_g) (x_f - x_g) - \lambda \frac{JT(q-1)}{2^{q-3}}}{g''(x_g)} + (x_f - x_g)h(x_g) + O(\lambda^2) \\ &\approx 1 - \frac{\lambda }{g''(x_g)}\frac{JT(q-1)}{2^{q-3}}  \\ &= 1 - \lambda \frac{JT(q-1)}{2^{q-1}}\sech^2\left({\frac{JT}{2^{q-1}}}\right) .
	\end{split}\label{eq:Hdet}
\end{equation}
In a similar manner we can evaluate the expression in the exponential. Finally, we get:
\begin{equation}
	F(T) \approx \left(1 + \lambda \frac{JT(q-1)}{2^{q}}\sech^2\left( \frac{JT}{2^{q-1}} \right)\right) \exp\left[\frac{N_1 \lambda JT(q-1)} {2^q}\exp\left(-\frac{JT}{2^{q-2}}\right)\sech^2\frac{JT}{2^q} \right] .
\end{equation}
The term in the exponential turns out to be equal to the $O(\lambda^2)$ contribution from the classical action while the factor multiplying the exponential piece is the contribution from the one-loop determinant that we computed in \eqref{eq:one-loop-calc}.

\section{Proof of the error correction bound}
\label{appQEC}

We use the two different measures of distance in the proof \cite{NielsenChuang}: the trace distance and fidelity.
The trace distance between two states $\rho$ and $\sigma$ is:
\begin{equation}
\begin{aligned}
    D(\rho,\sigma) &= \frac{1}{2} \Tr( | \rho - \sigma | )  \\ 
    &= \max_{Q} \Tr \left( Q (\rho - \sigma) \right) ,
    \end{aligned}
\end{equation}
where $|A| = \sqrt{A^\dagger A}$. In the second expression, we maximize over all possible projectors $Q$. 
The fidelity between two states $\rho$ and $\sigma$ is defined as:
\begin{equation} 
\begin{aligned}
F\left( \rho,\sigma \right) &= \Tr \left( \sqrt{\sqrt{\sigma} \rho \sqrt{\sigma}} \right) \\
&= \max_{\ket{\psi_\sigma}} | \inner{\psi_\rho}{\psi_\sigma}  | ,
\end{aligned}
\end{equation} 
where $\ket{\psi_\rho}$ and $\ket{\psi_\sigma}$ are purification of $\rho$ and $\sigma$ respectively. 

Consider a maximally entangled state $\ket{\Psi}$ between the encoded code subspace and a reference system isomorphic to the code subspace: 
\begin{equation}
    \ket{\Psi}= \sum_{i} \frac{1}{\sqrt{d_\co}} \ket{i}_{\re} \otimes \ket{\psi_i}_\ph .
 \end{equation}
The physical system interacts with the environment initially in some pure state $\ket{0}_{\en}$. This interaction is described by a joint evolution of the physical system and the environment by a unitary $U_\mathcal{E}$ leading to the following final state:
\begin{equation}
    \ket{\Psi'} = \sum_{i}\frac{1}{\sqrt{d_\co}}  \ket{i}_{\re} \otimes U_{\mathcal{E}} \left( \ket{\psi_i}_\ph \otimes \ket{0}_{\en}\right) .
\end{equation}
Consider now a fictitious state
\begin{equation}
    \tilde{\rho}_{\re,\en} = \rho'_\re \otimes \rho'_\en ,
\end{equation} where the reduced states are
\begin{equation}
\begin{aligned}
    \rho'_\re &= \Tr_{\en, \ph} \left( \ket{\Psi'}  
    \bra{\Psi'} \right), \\ 
    \rho'_\en &= \Tr_{\re, \ph} \left( \ket{\Psi'}  
    \bra{\Psi'} \right) . \\ 
    \end{aligned}
\end{equation}
Note that $\tilde{\rho}_{\re,\en}$ is \textit{not} the state $\rho'_{\re,\en}$, but is instead a factorized state between $\cH_\re$ and $\cH_\en$ that is built from its reduced states.
Consider a purification $\ket{\tilde{\Psi}}$ of $\tilde{\rho}_{\re,\en}$ such that its trace distance with $\ket{\Psi'}$, the quantity $D(\ket{\tilde{\Psi}}, \ket{\Psi'})$, is minimum. By the Schmidt decomposition of pure states, any purification of $\tilde{\rho}_{\re,\en}$ may be written in the following form:
\begin{equation}\label{eq:purifiedrefenv}
    \ket{\tilde{\Psi}} = \sum_{i,j}\sqrt{\frac{\alpha_j}{d_\co} } \ket{i}_{\re} \otimes  \ket{\phi_{ij}}_\ph \otimes \ket{j}_{\en}
\end{equation}
where the set $\{\ket{\phi_{ij}}\}$ form an orthonormal basis of the physical Hilbert space.
The Schmidt coefficients $\alpha_j$ depend only on the environment index because $\rho'_\re$ is maximally mixed which restricts the form of the Schmidt coefficients in this manner.
Indeed, the state $\rho'_\en$ determines the real non-negative coefficients $\sqrt{\alpha_j}$ completely.
The condition that $\ket{\tilde{\Psi}}$ should be a purification with minimal $D(\ket{\tilde{\Psi}},\ket{\Psi'})$ is hidden in the basis vectors $\ket{\phi_{ij}}_\ph$.
Define projection operators $\Pi_j$ as:
\begin{equation}
    \Pi_{j} = \sum_{i} \ket{\phi_{ij}}\bra{\phi_{ij}} .
\end{equation}
These projectors satisfy the following relation:
\begin{equation}
    \Pi_{j}\Pi_k = \delta_{jk} \Pi_k .
\end{equation}
Moreover, every subspace corresponding to the projector  $\Pi_j$ is isomorphic to the code subspace i.e. for each $\Pi_j$, there is a unitary operator $U_j$ such that $U_j \Pi_j U_j^\dagger = \Pi_{\text{code}}$, where $\Pi_{\text{code}}$ is a projector onto the code subspace. 

Following \cite{Schumacher2001}, we construct a recovery channel $\tilde{\cR}$ which consists of the following two operations: 
\begin{enumerate}
    \item Measurement with some projection operator $\Pi_j$, and 
    \item Rotation of the resulting state by the unitary operator $U_j$ .
\end{enumerate}
Consider acting with $\tilde{\cR}$ on $\ket{\tilde{\Psi}}$.
Measurement of $\ket{\tilde{\Psi}}$ with $\Pi_j$ projects the state $\ket{\tilde{\Psi}}$ to the following state with probability $\alpha_j$:
\begin{equation}
   \ket{\tilde{\Psi}}_j = \sum_{i}\frac{1}{\sqrt{d_\co}}  \ket{i}_{\re} \otimes  \ket{\phi_{ij}}_\ph \otimes \ket{j}_{\en} .
\end{equation}
The unitary transformation $U_j$ acts on $\ket{\tilde{\Psi}_j}$ as
\begin{equation}
\begin{split}
   U_j \ket{\tilde{\Psi}}_j &= \sum_{i}\frac{1}{\sqrt{d_\co}}  \ket{i}_{\re} \otimes U_j \ket{\phi_{ij}}_\ph \otimes \ket{j}_{\en} \\ &= \sum_{i}\frac{1}{\sqrt{d_\co}}  \ket{i}_{\re} \otimes \ket{\psi_i}_{\ph} \otimes \ket{j}_{\en} \\ &= \ket{\Psi} \otimes \ket{j}_\en .
   \end{split} 
\end{equation}
Thus, $\tilde{\cR}$ acts on $\ket{\tilde{\Psi}}$ to give back the original state $\ket{\Psi}$ because the unitary $U_j$ acts to precisely rotate the basis $\ket{\phi_{ij}}$ via $U_j \ket{\phi_{ij}} = \ket{\psi_i}$. However, we are interested in recovery from the state $\ket{\Psi'}$, after the action of the error channel. We will now rephrase the condition for approximate recovery, derived in \cite{Schumacher2001} in terms of the trace distance between the recovered state and the initial state, using the mutual purity between the reference and the environment. We have the following bound on the trace distance between the state obtained by action of the recovery channel $\mathcal{\tilde{R}}$ on the actual state $\ket{\Psi'}$ and the initial state $\ket{\Psi}$.
\begin{equation}\label{eq:trbound}
    \begin{split}
         D \left( \tilde{\mathcal{R}}(\ket{\Psi'}\bra{\Psi'}),  \ket{\Psi}\bra{\Psi}\right) &=  D \left( \tilde{\mathcal{R}}(\ket{\Psi'}\bra{\Psi'}), \tilde{\mathcal{R}}( \ket{\tilde{\Psi}}\bra{\tilde{\Psi}} )\right) \\ 
        &\leq   D \left( \ket{\Psi'}, \ket{\tilde{\Psi}} \right) \\
        &= \sqrt{1 - | \inner{\Psi'} {\tilde{\Psi} } |^2 }\\
        & = \sqrt{1 - F^2(\rho'_{\re,\en}, \rho'_{\re}\otimes \rho'_{\en})} \\
        &\leq \sqrt{2 - 2F(\rho'_{\re,\en},\rho'_\re \otimes \rho'_\en)}\\
        &\leq \sqrt{2D(\rho'_{\re,\en}, \rho'_{\re}\otimes \rho'_{\en})} \\ 
        &\leq\sqrt{ d_{\re}\, d_{\en}\, \lambda_{\text{max}} } \\ 
        &\leq  \sqrt{d_{\re}\, d_{\en}}\, \left( \Tr \left( \rho'_{\re,\en} - \rho'_\re \otimes \rho'_\en \right)^2 \right)^{1/4}  \\
        &= \sqrt{ d_{\re}\, d_{\en}}\, \left( \Tr\left( \rho'^2_{\re,\en} - \rho'^2_{\re}\otimes \rho'^2_{\en} \right)\right)^{1/4} . 
    \end{split}
\end{equation}
In the second step, we used the monotonicity property of trace distance  with respect to the action of a channel (see chapter 9 of \cite{NielsenChuang}). The fourth step follows from the definition of fidelity and the fact that $\ket{\tilde{\Psi}}$ is a purification of $\tilde{\rho}_{\re,\en}$ that minimizes its trace distance with $\ket{\Psi'}$. The sixth step is a standard inequality between fidelity and trace distance \cite{NielsenChuang}.  As in the main text, $d_\re$ and $d_\en$ are respective dimensions of the reference and the environment Hilbert spaces.  In the seventh step, $\lambda_{\text{max}}$ is the maximum eigenvalue of $|\rho'_{\re,\en} - \rho'_{\re}\otimes \rho'_\en|$. Since $\lambda^2_{\text{max}} < \Tr\,(\rho'_{\re,\en} - \rho'_\re\otimes \rho'_\en)^2$, the eighth step follows. The final step is true because $\rho'_\re$ is maximally mixed.

To summarize, we have shown that there exists a set of projection operators ${\Pi_j}$, the measurement of which followed by a unitary transformation with $U_j$ approximately recovers the maximally entangled state between the reference and the physical system.
The accuracy of this recovery in terms of trace distance is bounded by the combination we have found, which is the mutual purity $\cF_{\Psi'}(\re : \en)$ from the main text.

The inequality (\ref{eq:trbound}) was derived for a specific recovery channel $\tilde{\mathcal{R}}$. However,  there may exist a better recovery channel $\mathcal{R}$ which must also satisfy the inequality:
\begin{equation}\label{eq:bound}
    D \left( \mathcal{R}(\ket{\Psi'}\bra{\Psi'}),  \ket{\Psi}\bra{\Psi}\right) \leq \sqrt{ d_{\re}\, d_{\en}}\, \left( \Tr\left( \rho'^2_{\re,\en} - \rho'^2_{\re}\otimes \rho'^2_{\en} \right)\right)^{1/4} .
\end{equation}
 We can use the above inequality to compute a bound on recovery of arbitrary states in the code subspace after the action of the error channel $\mathcal{E}$. We will use the channel-state isomorphism of  \cite{PhysRevA.60.1888} as follows: Consider a state $\sigma = \sum_{m,n} \sigma_{mn} \ket{\psi_m} \bra{\psi_n}$ in the code subspace and let $\sigma' = \rerr{\sigma}$. We can write $\sigma'$ in terms of $ \sigma_{\re} =\sum_{m,n} \sigma_{mn} (\ket{m}\bra{n})_{\re}$ and $\omega = \rerr{\ket{\Psi} \bra{\Psi}}$ as
\begin{equation}
\begin{aligned}
    \sigma' &= \sum_{m,n} \sigma_{mn}\rerr{\ket{\psi_m}\bra{\psi_n} } \\
    &= d_{\re}  \sum_{k,l} \Tr_{\re} \left( \frac{1}{d_\re}\ket{k}\bra{l} \sigma^T_{\re} \right) \,  \rerr{\ket{\psi_k}\bra{\psi_l}} \\
    &= d_{\re} \Tr_{\re} \left( \sigma^T_{\re} \, \rerr{\ket{\Psi} \bra{\Psi}} \right) \, .
    \end{aligned}
\end{equation} 
Here $\sigma^T_\re$ is the transpose of $\sigma_\re$.
We have a similar expression for $\sigma$: \begin{equation}
\sigma =d_{\re} \Tr_{\re} \left( \sigma^T_{\re} \ket{\Psi}\bra{\Psi} \right) .
\end{equation}
We can derive a bound on the trace distance between $\sigma$ and $\sigma'$ as follows: 
\begin{equation}
\begin{split}
    D(\sigma',\sigma) &= \frac{1}{2} \,\Tr_{\ph}  | \sigma' - \sigma|  \\
    &= \frac{1}{2} \, d_\re \Tr_{\ph} | \Tr_{\re} \left( \sigma^T_{\re}  \left( \rerr{\ket{\Psi} \bra{\Psi}} - \ket{\Psi}\bra{\Psi} \right)\right) |  \\ 
   &\leq  \frac{1}{2} \,d_\re \Tr (| \sigma^T_\re (\rerr{\ket{\Psi} \bra{\Psi}} - \ket{\Psi} \bra{\Psi})|) \\
   &=  \frac{1}{2} \,d_\re \Tr (| \sigma^T_\re \otimes \Pi_{\text{code}} (\rerr{\ket{\Psi} \bra{\Psi}}- \ket{\Psi} \bra{\Psi})|) \\
   &\leq  d_\re \Tr\left( \sigma^T_\re\otimes \Pi_{\text{code}}\right)\, \frac{1}{2}  \Tr\left( |\rerr{\ket{\Psi} \bra{\Psi}} - \ket{\Psi} \bra{\Psi}|\right) \\ 
   &= d_\re \Tr\left( \sigma^T_\re\otimes \Pi_{\text{code}}\right)  D\left( \rerr{\ket{\Psi} \bra{\Psi}}, \ket{\Psi} \bra{\Psi}\right)\\
     &\leq d^{5/2}_{\re}\, d_\en^{1/2}\left( \Tr\left( \rho'^2_{\re,\en} - \rho'^2_{\re} \otimes \rho'^2_{\en} \right)\right)^{1/4}  .
    \end{split}
\end{equation}
In the fifth step, $\Pi_{\text{code}}$  is the projector on the code subspace. In the final step, we used the inequality in (\ref{eq:bound}).
The result above is precisely the one quoted in \eqref{eq:Renyi2Bound}.

\section*{} 
\bibliographystyle{JHEP}
\bibliography{refs}

\end{document}